\documentclass[12pt]{article}

\usepackage[utf8]{inputenc}
\usepackage{charter}
\usepackage{fullpage}
\usepackage{hyperref}
\usepackage{graphicx}
\usepackage{lipsum}
\usepackage[sort&compress,numbers]{natbib}
\usepackage{hyperref}
\hypersetup{hidelinks}
\usepackage{amsmath}
\usepackage{amssymb, xspace}
\usepackage[total={6.5in,9in},top=1in,headsep=0.1in,headheight=1in]{geometry}
\usepackage{siunitx} 
\usepackage{aas_macros}
\usepackage{multirow}

\newcommand\snowmass{
\begin{center}
  \rule[-0.2in]{\hsize}{0.01in}\\
  \rule{\hsize}{0.01in}\\
  \vskip 0.1in
  Submitted to the Proceedings of the US Community Study\\ 
  on the Future of Particle Physics (Snowmass 2021)\\
  \rule{\hsize}{0.01in}\\
  \rule[+0.2in]{\hsize}{0.01in}\\[-2em]
\end{center}
}
\newcommand{\tcm}{21$\,$cm\xspace}  
\newcommand{\aprx}{\mbox{$\ensuremath{\sim}$}}
\newcommand{\Tspin}{\ensuremath{T_s}}
\newcommand{\Tcmb}{\ensuremath{T_\gamma}}
\newcommand{\nuHI}{\ensuremath{\nu_{\scriptscriptstyle 21}}}

\newcommand{\sHI}{\ensuremath{{\scriptscriptstyle HI}}}

\newcommand{\OmegaHI}{\ensuremath{\Omega_\sHI}}

\usepackage[firstpage=true]{background}
\backgroundsetup{contents={\parbox{6.5in}{\snowmass}}, scale=1,placement=top,opacity=1,color=black,position={3.25in,1.2in}}

\usepackage{fancyhdr}
\fancypagestyle{plain}{%
  \fancyhf{}%
  \fancyhead[C]{}
  \fancyfoot[C]{\thepage}
}

\fancypagestyle{empty}{%
  \fancyhf{}%
  \fancyhead[C]{{\it Snowmass2021 Cosmic Frontier White Paper Template}}
  \fancyfoot[C]{\thepage}
}
\title{Snowmass2021 Cosmic Frontier White Paper: 21\,cm Radiation as a Probe of Physics Across Cosmic Ages}
\date{}

\usepackage{authblk}

\author[1]{Adrian Liu}
\author[2]{Laura Newburgh}
\author[3]{Benjamin Saliwanchik}
\author[4]{An\v{z}e Slosar}
\author[\space]{\\ for the Snowmass 2021 Cosmic Frontier 5 Topical Group}

\affil[1]{Department of Physics, McGill University, Montréal QC, Canada}
\affil[2]{Department of Physics, Yale University, New Haven CT 06520, USA}
\affil[3]{Instrumentation Division, Brookhaven National Laboratory, Upton NY 11973, USA}
\affil[4]{Physics Department, Brookhaven National Laboratory, Upton NY 11973, USA}

\begin{document}

\maketitle

\begin{abstract}
The 21\,cm line refers to a forbidden transition in neutral hydrogen associated with alignment of spins of the proton and electron. It is a very low energy transition that is emitted whenever there is neutral hydrogen in the Universe. Since baryons are mostly ($\sim$ 75\%) hydrogen, one can in principle detect this emission throughout much of the history of the Universe. The dominant emission mechanism is different across cosmic ages. Before the photons decouple from matter, hydrogen is in an ionized state and does not emit in 21\, cm. After recombination and during the Dark Ages, at  $z\sim 30 -1000$, the 21\,cm emission is associated with density fluctuations in the neutral hydrogen medium. After the first stars turn on and galaxies begin to form, the 21\,cm emission traces bubbles of ionized hydrogen in the sea of the neutral medium. This epoch, spanning $z\sim 6-30$, is often referred to as cosmic dawn and the Epoch of Reionization (EoR). At redshifts below $z<6$, the intergalactic medium is largely ionized, but pockets of self-shielded neutral gas form in dense galactic environments and 21\,cm emission traces the distribution of galaxies. The vastly different emission mechanisms allow us to probe very different physics at different redshifts, corresponding to different observational frequencies. The instrumental challenges, namely building very sensitive and exquisitely calibrated radio telescopes, however, share many commonalities across frequency bands. The potential of the 21\,cm probe has been recognized by the Decadal Survey of Astronomy \& Astrophysics, whose Panel on Cosmology identified the Dark Ages as its sole discovery area. We argue that HEP should recognize the potential of 21\,cm as a probe of fundamental physics across many axes and invest in the technology development that will enable full exploitation of this rich technique.
\end{abstract}

\newpage

\section*{Executive Summary}

Neutral hydrogen in the Universe emits radiation at the 21\,cm rest wavelength across cosmic ages after the epoch of recombination. This radiation is emitted by different dominating mechanisms across the history of the Universe and can probe numerous questions regarding the fundamental physics of the Universe and its constituents. In particular:
\begin{itemize}
    \item The Dark Ages monopole sensitively probes any source of injection or sink of energy during the epoch of the Universe between recombination and the appearance of the first stars. It can uniquely constrain models of dark matter and non-standard atomic physics over this period of time. Measurement of the monopole frequency spectrum is achievable over the next decade, establishing the necessary groundwork towards measuring density fluctuations during the Dark Ages. See also Snowmass 2021 Dark Matter facilities White Paper \cite{DMfacilitiesWP}.
    
    \item Fluctuations in the Dark Ages would open up the most constraining probe of cosmology known. The large, pristine, three-dimensional field contains orders of magnitude more information that the cosmic microwave background. However, it is technically very difficult and multiple decades away. The Decadal panel on Astronomy and Astrophysics 2020 has identified Dark Ages cosmology as the sole discovery area.
    
    \item The Epoch of Reionization (EoR) probes non-standard stellar evolution in the early Universe. It is sensitive to new interactions in baryons and dark matter physics. The large contrast of the ionized bubbles in the sea of neutral matter makes it observationally particularly appealing. EoR instrumentation is among the most advanced in radio astronomy. The detection of the signal is expected in the coming decade.
    
    \item Low redshift 21\,cm cosmology measures the neutral hydrogen after reionization in individual galaxies through their aggregate signal. It is sensitive to fluctuations in the galaxy density field at redshift $z<6$. It offers potentially the most sensitive probe of large-scale fluctuations in the Universe as outlined in the White Papers \cite{CF4_NLin} and \cite{CF5_inflation}.
\end{itemize}

We urge the P5 to consider the following recommendations for development of 21\,cm cosmology:

\begin{enumerate}
    
    \item DOE should continue to invest in the Dark Ages Lunar program, following the exploratory LuSEE-Night pathfinder program using synergies offered by NASA's Commercial Lunar Payload Services (CLPS).
    
    \item The community should invest in detection of the Dark Ages monopole, a probe of fundamental physics, from space, in lunar orbit, or from the lunar surface over the next decade in collaboration between DOE and NASA.
    
    \item The community should invest in the relatively inexpensive program to robustly detect the reionization signature in the radio monopole from the ground in the next decade.
    
    \item DOE should invest in a long-term advanced detector program to develop technologies that will enable robust 21\,cm physics measurements across the full range of radio frequencies: from low-power DAQs used in the lunar environment, to high bandwidth RF System-on-Chip based solutions for proposed experiments such as PUM as well as general development of methods that advance the calibration precision.  This effort should be viewed as a long-term strategic development towards enabling new cosmic frontier programs in the future.
    
    \item DOE should invest in development of simulation and data analysis techniques to leverage the full richness of the data from 21\,cm cosmology. This development should focus on large computer simulations allowing end-to-end understanding of synthetic data and enabling development of precision algorithms and data reduction codes. All these requirements are an excellent match to DOE institutional strengths.
    
\end{enumerate}

\newpage

\section{Introduction}

This white-paper focuses on the spin-flip transition in neutral hydrogen as a probe of fundamental physics of the Universe. In other white-papers, this technique has been discussed through different possible intersections in the space of theories and experiments. For example, in the  Snowmass2021 White Paper \cite{DMfacilitiesWP} the technique is discussed through the lens of its potential to constrain the dark matter, while in the Snowmass2021 White Paper \cite{CF4_NLin} its potential is viewed from perspective of measuring structure in the Universe. In this Cosmic Frontier 5 coordinated White Paper we focus on commonalities between 21\,cm measurements at different frequencies and focus on its potential to open new avenues for studying fundamental physics through observations of the Universe.

Our main message could be summarized as follows: observations of 21\,cm across cosmic ages offers a unique insight into different aspects of fundamental physics. The physics of emissions varies from epoch to epoch and therefore the sensitivity to different aspects of fundamental physics vary. Still, the technical steps required to be able to extract this information have many commonalities. \textbf{By investing in a coordinated long-term strategic 21\,cm development program, the DOE could reap synergistic benefits across the wide range of physics probed by different frequencies.}

We have based the contents of this white-paper on various resources that were already written by the authors, including references \cite{2018arXiv181009572C,2020PASP..132f2001L}.

This paper is structured as follows. In Section \ref{sec:theory} we introduce various mechanism of 21\,cm emission and discuss some of the opportunities for the physics which can be probed. In Section \ref{sec:challenges} we describe the challenges that have prevented us from achieving this so far. We stress that these challenges are purely technical in nature. In Section \ref{sec:facilities} we discuss the currently funded and planned facilities. We conclude with a concrete set of recommendations for consideration by the P5 panel.


\section{21\,cm emission Across Cosmic Ages}
\label{sec:theory}

The \tcm line is the hyperfine transition of atomic hydrogen. The parallel alignment of the electron and proton spins is a slightly higher energy state than the anti-parallel alignment. As an atom transitions from one state to the other, it emits (or absorbs) a photon of \tcm wavelength. To study this line we use a quantity called the spin-temperature $\Tspin$ that describes the relative occupancy of the two spin states.
\begin{equation}
\label{eq:SpinTemperature}
    \frac{n_1}{n_0} = 3 \: \exp{\biggl(- \frac{h \, \nuHI}{k_b \, \Tspin}\biggr)} \; ,
\end{equation}
where the factor of three comes from the relative degeneracy of the states, $n_1$ is the number of atoms in the excited hyperfine state, $n_0$ is the number in the ground hyperfine state, $h$ is Planck's constant, $k_b$ is Boltzmann's constant, and $\nuHI \approx 1420.406\,\textrm{MHz}$ is the rest frequency of the 21\,cm line. The physics of the spin temperature throughout the history of the Universe is complex; we will cover the relevant parts below. The key thing to note is that we observe the \emph{contrast} between the Cosmic Microwave Background (CMB) and the spin temperature. Where the spin temperature is higher than the CMB temperature, $\Tcmb$, we are emitting photons and see an excess above the CMB temperature; when it is lower than $\Tcmb$ the photons are absorbed from the CMB and we see a deficit compared to what we expect.\footnote{In this paper, we focus exclusively on absorption or emission of the 21\,cm line relative to the CMB. However, it may also be possible to detect 21\,cm absorption from a bright, high-redshift radio source \citep{ChrisRadioLoud2002,FurlanettoRadioLoud2002,XuRadioLoud2009,neosporin,CiardiRadioLoud2013,EwallWiceRadioLoud2014,CiardiRadioLoud2015,Ciardi2015RadioLoudSKA,Smelin2016}.}

The brightness temperature $T_b$ of the 21\,cm line is given by
\begin{equation}
\label{eq:Tb_theory}
T_b(\mathbf{\hat{r}}, \nu) = \left(\frac{3 \hbar c^3 A_{10} }{16 k_b \nu_{_{21}}^2 } \right) \left[\frac{ x_\textrm{HI} n_\textrm{H}}{(1 + z)^2 (dv_\parallel / dr_\parallel) }  \right] \left( 1-\frac{T_\gamma  }{T_s} \right)
\end{equation}
One thing to note is that in the high spin temperature limit the observed brightness temperature is independent of the spin temperature itself. This can be understood from the fact that at high temperatures all spin microstates are equally occupied and thus the observed brightness depends only on the rate of spontaneous emission rate from the high energy state ($A_{10}$).

Regardless of the type of observation, one sees from Equation \eqref{eq:Tb_theory} that the 21\,cm brightness temperature contains a rich variety of effects. For example, at certain redshifts $T_s$ is strongly coupled to the baryonic gas temperature (see sections below). This makes $T_b$ a high-redshift thermometer (albeit a rather indirect one). It is also a probe of the ionization state of hydrogen, via its dependence on $x_{\rm HI}$. We therefore see that $T_b$ is likely to be an excellent probe of many high-redshift astrophysical processes. In addition, it is sensitive to cosmology, since the distribution of hydrogen (entering via the factor of $n_H$) is driven by the large scale cosmological distribution of matter, as is the $dv_\parallel / dr_\parallel$ velocity term, since it includes peculiar velocities in addition to the Hubble flow. Of course, with all of the aforementioned effects contributing to $T_b$, it may be difficult to cleanly probe any of them. Fortunately, different phenomena tend to dominate at different redshifts, and in what follows we provide a quick qualitative description of this.

\begin{figure*}[ht!]
\centering
\includegraphics[width=1.00\textwidth,trim={0cm 0cm 0cm 0cm},clip]{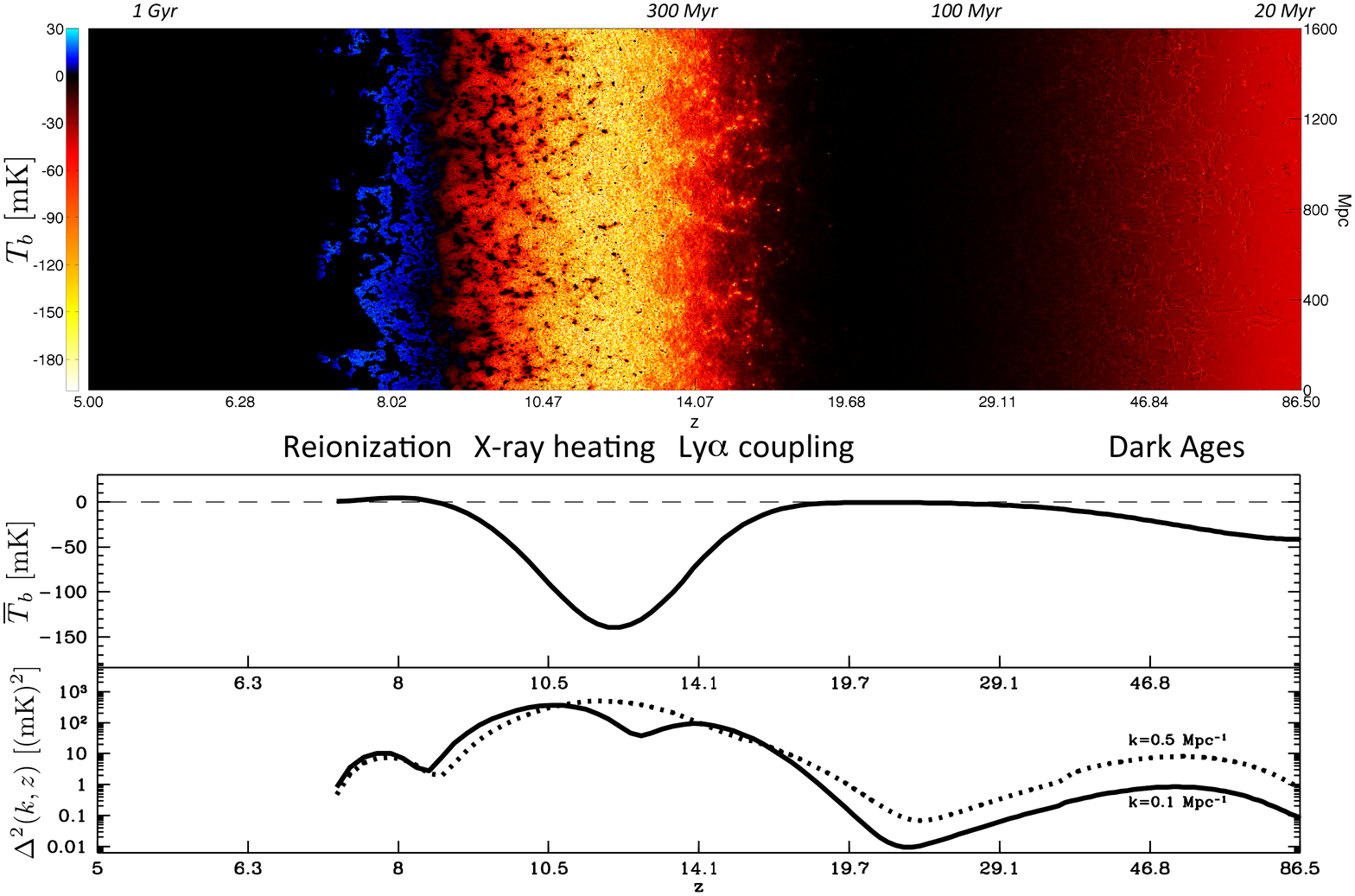}
\caption{A graphical representation of a simulated 21\,cm signal from $z \sim 90$ (during the Dark Ages) to the $z \sim 7$ (the end of reionization in this particular simulation). The top panel shows the 21\,cm brightness temperature contrast with the CMB, displayed as a two-dimensional slice of a three-dimensional volume. Because the line-of-sight distance for the 21\,cm line is obtained via the redshift, the horizontal axis also doubles as a redshift/evolution axis (i.e., the top panel is a picture of a \emph{light cone}, which is what a real telescope would observe). The middle panel shows the predicted \emph{global signal} of the 21\,cm line, where the spatial information in the vertical direction of the top panel has been averaged over to produce one average brightness temperature per redshift. The bottom panel shows the power spectrum, expressed as $\Delta^2 (k) \equiv k^3 P(k) / 2 \pi^2$, as a function of redshift. This can be thought of as a measure of the variance of spatial fluctuations as a function of spatial wavenumber $k$, with the solid line showing $k = 0.1\,\textrm{Mpc}^{-1}$ and the dotted line showing $k=0.5\,\textrm{Mpc}^{-1}$). These simulations were generated from the \texttt{21cmFAST} semi-analytic code (
\url{https://github.com/andreimesinger/21cmFAST}), and are publicly available at \url{http://homepage.sns.it/mesinger/EOS.html} as part of the Evolution of 21\,cm Structure project \citep{Mesinger2016EOS}.}
\label{fig:EOS}
\end{figure*}

\subsection{Ultra high redshift: the Dark Ages}

After recombination, though the photons and baryons no longer act as a single fluid, scatterings between photons and residual electrons (and collisions with the rest of the gas) mean that the baryon temperature is coupled to the photon temperature until $z \sim 300$ when these electron-photon scatterings become rare. At this point, the gas starts to cool faster than the photons $\propto (1 + z)^2$, and because the collisional coupling between the gas temperature and the spin temperature remains strong, Equation \eqref{eq:Tb_theory} predicts a net 21\,cm absorption signal. All of this takes place during the \emph{Dark Ages}, a pristine cosmological era where the first luminous sources have yet to form \citep{Scott1990}, rendering the 21\,cm line a direct probe of density fluctuations (at least to the extent that hydrogen traces the overall matter distribution). This era stops being observationally accessible at $z \sim 30$, when collisions between neutral hydrogen atoms become rare enough that the spin degrees of freedom are no longer coupled to the kinetic degrees of freedom, and the spin temperature rises until it equilibrates with the CMB temperature. This is illustrated in the middle panel of Figure \ref{fig:EOS}, where we show the global 21\,cm signal. With the CMB temperature and the spin temperature in equilibrium, the brightness temperature contrast goes to zero and there is no 21\,cm signal to observe.

The Dark Ages are of tremendous cosmological interest. The epoch provides access to a huge number of Fourier modes of the matter density fields \citep{Loeb2004}. These fluctuations are graphically depicted in the red high redshift regions of the top panel of Figure \ref{fig:EOS}. They are a particularly clean probe of the matter field since this is prior to the formation of the first luminous sources, thus obviating the need for the modelling of complicated astrophysics. The theoretical modelling effort is in fact even simpler than, say, that needed for galaxy surveys, since at these high redshifts, matter fluctuations are in the linear perturbative regime even at very fine scales. This is important because the spatial fluctuations exist to extremely small scales, as they are not Silk damped and therefore persist down to the Jeans scale \citep{Tegmark2009}. The result is a vast number of modes that can be in principle be used to probe fundamental cosmology. The long lever arm in scale, from the largest structures to the smallest structures, enables constraints on the running of the matter power spectrum's spectral index \citep{Mao2008}. This would represent an incisive probe of the inflationary paradigm. Future measurements may also be able to detect features in the primordial power spectrum \citep{Chen2016Features}, or to detect primordial non-Gaussianity~\citep{Munoz2015nonGaussianity} (see also~\citep{CF5_inflation}). In addition, the 21\,cm line can be used to probe the existence of relic gravitational waves from inflation~\citep{CF5_inflation}, either through their direct effects on large scale structure \citep{Masui2010} or through their lensing effects \citep{Book2012}. Tests of statistical isotropy and homogeneity may also be possible \citep{Shiraishi2016}. Small-scale measurements enable measurements of the neutrino mass \citep{Mao2008} and constraints on the existence of warm dark matter \citep{Loeb2004}. Finally, the cleanliness of measurements during the Dark Ages provide a good platform for detecting exotic phenomena beyond the standard models of particle physics and cosmology (see references in \cite{Furlanetto2019fundamentalphysics}).

Measurements during this era are extremely challenging. Not only do the extremely long wavelengths ($\lambda \approx 7$ to $70\,\textrm{m}$) make simply building an instrument with the required resolution and sensitivity extremely difficult, but as we will discuss later, the foregrounds are extremely bright. Additionally, the ionosphere is opaque at frequencies lower than a few MHz, and can cause distortions even at higher frequencies forcing observations to be done from space, where the far side of the moon currently holds most promise as the observatory.

\subsection{High redshift: Cosmic Dawn and the Epoch of Reionization}

As the first luminous objects begin to form, the 21\,cm line ceases to directly trace the dark matter distribution. Radiation from these objects affects the spin temperature and ionization state of the intergalactic medium (IGM) in a spatially and temporally non-trivial way. This is reflected in the spatial fluctuations and redshift evolution of the 21\,cm line, which is no longer governed by the relatively simple physics of cosmological matter perturbations alone. This is both a defect and an opportunity. It is an opportunity because it provides a promising tool for understanding the nature of the first luminous objects. It is a defect because the complicated astrophysics of the era (often loosely\footnote{There is unfortunately no consistent definition of Cosmic Dawn that is agreed upon in the literature. Some authors use it to refer to a period that began when the first stars formed, and ended with the formation of larger galaxies. Others define the end of Cosmic Dawn to be when the first galaxies began to systematically reionize the intergalactic medium. Yet others consider Cosmic Dawn to be a broad term that encompasses the entire period from the formation of the first stars to the end of reionization.} referred to as ``Cosmic Dawn") means that one must be more creative in extracting information about fundamental physics. However, several promising avenues have emerged in recent years on this front. For example, velocity-induced acoustic oscillations \citep{Dalal2010,Fialkov2012VAOs,Munoz2019VAOs1,McQuinn2012VAOs} have the potential to serve as clean standard rulers at high redshifts, thus enabling precision measurements of the Hubble expansion rate \citep{Munoz2019VAOs2}. It may also be possible to parametrize (and marginalize out) any messy astrophysics in a fairly model-agnostic way using effective field theory techniques \cite{McQuinn:2018, Hoffmann:2019}. Machine-learning techniques may also allow the ``undoing" of some astrophysical effects \cite{2021ApJ...907...44V}. Probes of small-scale cosmological structure through their couplings to large scales have shown promising signs of being relatively robust to changes in astrophysical modelling \cite{2020PhRvD.101f3526M}. Finally, combinations of 21\,cm measurements with CMB measurements can provide complementary constraints that remove reionization as a nuisance for CMB measurements \cite{2021arXiv211206933B}, sharpening constraints on fundamental physics \cite{Liu:2016,Billings:2021,Fialkov:2016}.

\begin{figure}[t]
\centering
\includegraphics[width=0.45\textwidth,trim={3cm 0.75cm 2cm 1cm},clip]{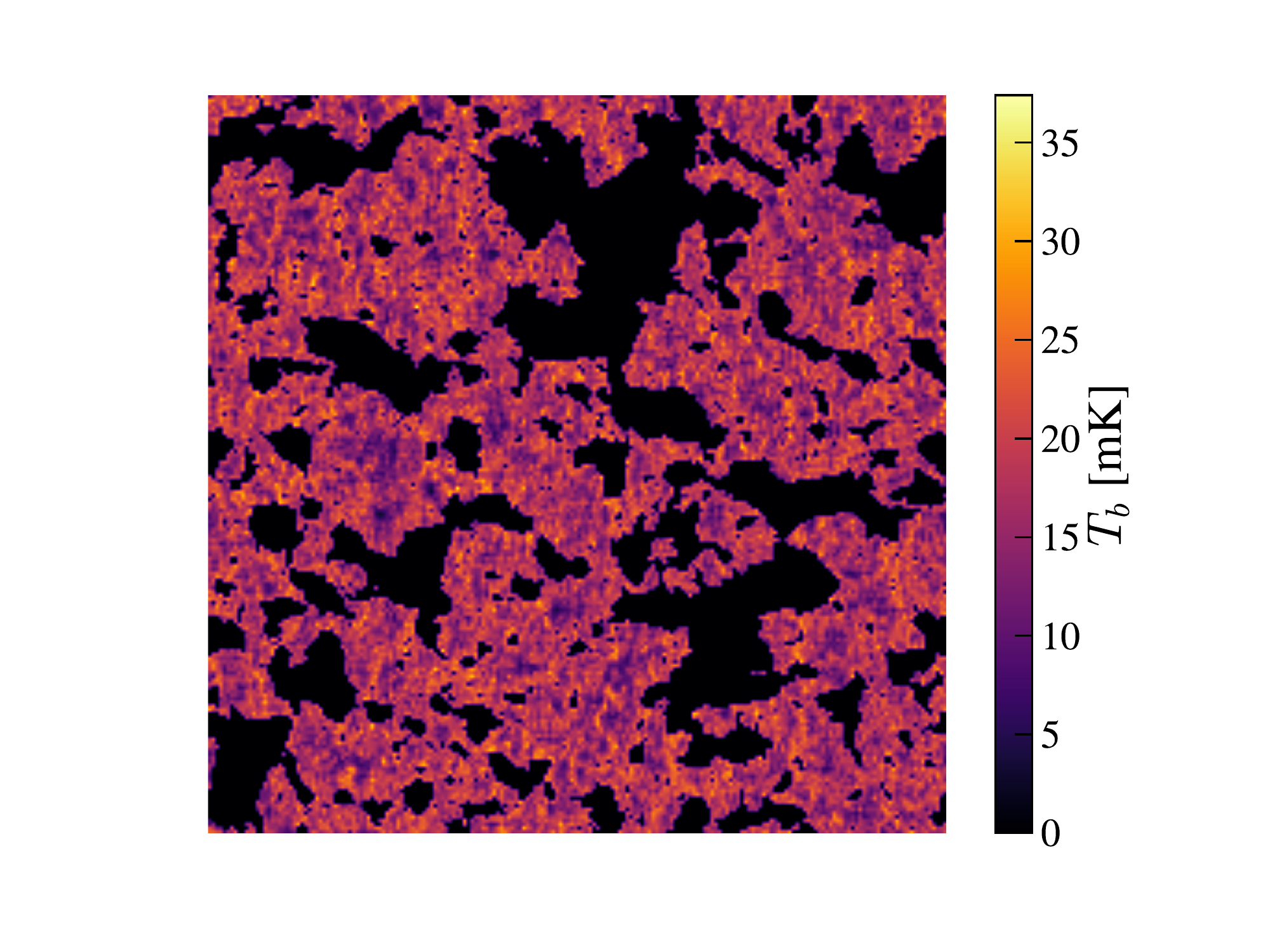}
\caption{An example 21\,cm brightness temperature field at $z=8.1$, generated using the \texttt{21cmFAST} semi-analytic code \citep{Mesinger2011}. With the particular astrophysical parameters and models used to generate this simulation, $z=8.1$ corresponds to a volume-weighted neutral fraction of $\sim 0.55$. Ionized regions are shown in black, with zero brightness temperature since there is no 21\,cm transition when there is no neutral hydrogen present. The spatial fluctuations shown here encode a wealth of information about both astrophysics and cosmology, which 21\,cm experiments aim to extract.}
\label{fig:21cmFAST}
\end{figure}

Figure \ref{fig:EOS} illustrates a series of three epochs following the Dark Ages:
\begin{enumerate}
\item Cosmic Dawn begins with a period of \textbf{Lyman alpha coupling}, which runs from $z \sim 20$ to $z \sim 12$ for the model shown in Figure \ref{fig:21cmFAST}. As the first stars are formed, they produce significant amounts of Ly$\alpha$ flux. This causes the Wouthuysen-Field effect, whereby Ly$\alpha$ photon absorption promotes an electron in a neutral hydrogen atom from an $n=1$ state to an $n=2$ state, only to be followed by a decay to a \emph{different} hyperfine state when the electron returns to the $n=1$ state. This enables Ly$\alpha$ photons to induce 21\,cm spin-flip transitions, and because of the large cross-section of Ly$\alpha$ scattering, this causes the spin temperature $T_s$ to be coupled to the gas temperature. As was the case during the Dark Ages, the gas temperature is cooler than $\Tcmb$ in this epoch, which in turn means that the spin temperature $T_s$ must also be cooler than $\Tcmb$. The result is an absorption signal that contains information about both the density field and the Ly$\alpha$ flux. As star formation continues, the Ly$\alpha$ background eventually becomes sufficiently strong for the coupling between Ly$\alpha$ photons and gas kinetics to be extremely efficient everywhere. Fluctuations from the Ly$\alpha$ background then become negligible.
\item With continued star formation (and eventually galaxy assembly), we enter a period of \textbf{X-ray heating}. High-energy photons from the first luminous objects have a heating effect on the IGM, as these photons cause photo-ionizations of HI and HeI, with the resulting photoelectrons colliding with other particles. This can result in heating, further ionizations, or atomic excitations. The heating contribution raises the gas temperature to be above $\Tcmb$, which makes the 21\,cm signal go from absorption to emission. Because the astrophysical sources that heat the IGM are spatially clustered, this results in spatial fluctuations in the 21\,cm signal. Eventually, however, the entire IGM is heated and the fluctuations sourced by X-ray heating disappear.
\item Cosmic Dawn ends with the \textbf{Epoch of Reionization (EoR)}, when sustained star formation provides enough ionizing photons to systematically ionize the IGM. Reionization does not proceed uniformly because the objects responsible for producing the ionizing photons (likely galaxies) are clustered. This is illustrated in Figure \ref{fig:21cmFAST}, where ionized bubbles form and grow around regions of high matter density because it is those regions that contain the most galaxies (and therefore produce the most ionizing photons). Since the ionized regions contain almost no neutral hydrogen,\footnote{The ``ionized" regions shown in Figure \ref{fig:21cmFAST} are not entirely ionized. At smaller scales  (which are unresolved in the simulation), there exist galaxies and neutral gas clouds that are sufficiently dense to be self-shielded from ionization \citep{Sobacchi2014,Watkinson2015selfshielding}.} the 21\,cm brightness temperature is close to zero there. The intricate pattern of ionized versus neutral regions gives rise to strong spatial fluctuations in the 21\,cm line. These fluctuations persist until the ionized bubbles are sufficiently large and numerous that they overlap, and reionization of the entire IGM is complete.
\end{enumerate}

From the bottom panel of Figure \ref{fig:EOS} (which shows the fluctuation power as a function of redshift) we see that each of the three epochs shows a rise-and-fall pattern: the fluctuations increase when a source of inhomogeneity (Ly$\alpha$ coupling, X-ray heating, or ionization) becomes relevant, peak when (loosely speaking) roughly half of the volume is affected, and fall once the relevant process has uniformly affected our Universe. This provides a set of distinctive signatures to search for in observational data \citep{Lidz2008RiseAndFall,Christian2013}. However, it is important to stress that even if our qualitative description of Cosmic Dawn is accurate, there still remains a wide landscape of possible 21\,cm predictions.


Fortunately, direct observations of the 21\,cm line at the redshifts of Cosmic Dawn should significantly enhance our understanding. For example, the typical sizes of ionized bubbles will place constraints on the nature of sources responsible for reionization \citep{McQuinn2007}. A single 21\,cm measurement of the midpoint of reionization (performed, perhaps, by locating the peak of ionization fluctuation power) would inform the entire timeline of reionization when combined with CMB and Ly$\alpha$ forest data \citep{Pritchard2010BayesianCombo}. Detailed studies of 21\,cm fluctuations over a wide variety of spatial scales and redshift will enable constraints on the high-redshift Universe, shedding light on details such as the minimum mass for halos to host active star forming galaxies, or the X-ray luminosity of galaxies, among other parameters \citep{Pober2013,Mesinger2014,Fialkov2017Xrays,Greig2017,Kern2017,Park2019}. In addition to other possible constraints that we have not discussed above, direct 21\,cm measurements will also play the important role of allowing models of Cosmic Dawn to be tested, rather than assumed.

\subsection{Low redshift: Post-Reionization}
\label{sec:LowzHI}
At low redshift, after the end of reionization, the neutral fraction has been driven down to $x_\sHI \sim 2\%$ \citep{Villaescusa-Navarro2018,Planck_params2015}. The gas that remains is in dense regions that have managed to self-shield against the ionizing background. Using simulations as a guide we expect the remaining HI to be found within a range of halo masses $\sim 10^{10}$ to $10^{13}\,M_\odot$. At the lowest mass end halos are not dense enough to effectively self shield, and at the high mass end where we are considering clusters, tidal stripping and ram pressure remove gas from within the constituent galaxies \citep{Villaescusa-Navarro2018}. Although the HI that remains can be either part of the cold ($T \lesssim 100\,{\rm K}$) or warm ($T \gtrsim 5000\,{\rm K}$) neutral medium, both temperatures are warmer than the CMB temperature within this redshift range and we expect to see the 21\,cm line in emission.

This redshift range is the most probed observationally. At very low redshift HIPASS ($z < 0.03$, \cite{HIPASS}) and ALFALFA ($z < 0.06$, \cite{2018MNRAS.477....2J}) have detected individual extra galactic objects through their 21\,cm emission allowing constraints on the mass function and abundance of low redshift neutral hydrogen finding that $\OmegaHI \sim (3.9 \pm 0.6)\times 10^{-4}$ \citep{2018MNRAS.477....2J}. At higher redshifts cross correlations with optical tracers have allowed detection of 21\,cm emission. This was initially performed with the Green Bank Telescope combined with DEEP2 and later WiggleZ which estimated the hydrogen abundance (times the bias factor $b_\sHI$) at $z \sim 0.8$ to be $\OmegaHI \, b_\sHI = 6.2^{+2.3}_{-1.5} \times 10^{-4}$ \citep{Switzer2013}.

The measurements described above are all direct measurements of 21\,cm emission but at higher redshifts we can use measurements of Damped Lyman Alpha systems to constrain the HI abundance up to $z = 4.9$, finding $\OmegaHI = 9.8^{+2.0}_{-1.8} \times 10^{-4}$ at $z = 4.9$ \citep{2015MNRAS.452..217C}. Together this means that unlike other epochs there is a well defined target for the amplitude of the signal.


As we expect the HI to be hosted in generally quite low mass halos, and the fraction not bound within haloes to be small (10\% at $z=5$, smaller at low redshift) the neutral hydrogen is an excellent tracer of the total mass distribution \citep{Villaescusa-Navarro2018}. This makes low redshift 21\,cm observations an excellent probe for cosmology, allowing us to pursue the kinds of science that are currently done by galaxy redshift surveys, but within the radio band. However, unlike galaxy redshift surveys we do not need to resolve individual galaxies (which is required only to determine their redshift), and can instead map the unresolved emission of all HI at each frequency of observation, which we can map directly to a redshift. The angular resolution is chosen to match scales of scientific interest, and each observed voxel may contain hundreds to thousands of individual galaxies. This is the idea of Intensity Mapping \citep{2004MNRAS.355.1339B,Peterson:2006,Chang2008,2008MNRAS.383..606W}, hereafter referred to as IM.


Observations within this epoch are mapping the large-scale structure of the Universe, giving them the ability to do much of the same science as optical galaxy redshift surveys. However, as they can naturally have wide fields of view and observe all redshifts in their band simultaneously, IM experiments are more readily able to quickly survey very large volumes than optical galaxy redshift surveys. And as they do not need to resolve individual galaxies, they can more easily push to high redshifts, through the redshift desert ($z\sim 1$--$3$, where optical spectroscopy is challenging) and beyond ($3 \lesssim z \lesssim 6$).

As IM is a very recent technique that is still being developed, the initial target within this era are to measure Baryon Acoustic Oscillations (BAO). These are the remnants of primordial sound waves that leave a distinct signature in the correlation structure of matter, and they can be used as a standard ruler, albeit a statistical one, to measure the expansion history of the Universe \citep{1998ApJ...504L..57E,Seo2003}. The BAO signature is quite distinct and thus more robust to systematics, making it an ideal first science target for 21\,cm IM experiments. As a bonus, one can push to ranges in redshifts not currently probed by existing optical surveys \citep{Peterson:2006,Chang2008,2008MNRAS.383.1195W}. By constraining the expansion history of our Universe we hope to be able to infer the properties of Dark Energy, in particular its equation of state $w(z)$, which can yield clues to a microphysical explanation. Such constraints on the equation of state could potentially shed light on recent tensions in measurements of the Hubble parameter \citep{Knox2019HubbleHunter}, particularly in the context of proposed explanations that have non-trivial time evolution of dark energy at $z \gtrsim 1$ (e.g., \cite{DiValentino2016,DiValentino2017,Keeley2019}).

As IM is able to survey very large volumes of the Universe with precise radial distances (unlike photometric redshift surveys) it is ideal for discovering small statistical effects where measuring a large number of modes is essential. One target is looking for features in the power spectrum which might tell us about inflationary physics \citep{Chluba2018, Slosar:2019gvt, CF5_inflation}. Another is looking for signatures of new physics by searching for non-Gaussianity in 21\,cm data \citep{2017PhRvD..96f3525L, Meerburg:2019qqi, CF5_inflation}.

The expansion history information that IM can obtain from BAO, combined with broadband measurements of the power spectrum shape, plus future Cosmic Microwave Background measurements gives a potent combination for probing the contents of the Universe. In particular the number of relativistic degrees of freedom and the sum of the neutrino masses \citep{2016JCAP...02..008O,2018JCAP...05..004O} can be constrained to $\lesssim 20\,{\rm meV}$ by combining IM with other probes.

On extremely large scales there are general relativistic corrections to the standard observables \citep{2011PhRvD..84d3516C}, that if observed could be a stringent confirmation of the current cosmological model. To measure these, one needs to map large volumes of the Universe to build up enough samples of these scales (for which IM is ideally suited), and to combine with other probes on similar scales (e.g., photometric surveys like the Large Synoptic Survey Telescope) in order to remove sample variance effects \citep{2015PhRvD..92f3525A}.

Although one only directly measures the inhomogeneity in the distribution of HI with 21\,cm IM, by looking at the distortions in the observed three-dimensional field caused by gravitational lensing, one can in principle infer the \emph{total matter} distribution within the volume \citep{Foreman2018}. As the lensing displacements are typically small, upcoming 21\,cm experiments may only be able to see this effect in cross correlation with photometric redshift surveys, but it will be extremely powerful for following generations of instruments.

Finally, although most of the science targets we have outlined above are cosmological, there is tremendous astrophysical interest in the nature of HI at low redshift. IM by itself gives us direct access to $\Omega_\sHI(z)$, the total amount of neutral hydrogen across redshift, and by using cross-correlations against optical tracers we can start to obtain information about which galaxies host the HI \citep{2017MNRAS.470.3220W}.


\section{Challenges and opportunities}
\label{sec:challenges}

While $21\,$cm intensity mapping provides an efficient
means of measuring structure to high redshift, the instrument and analysis must be designed to overcome
systematic sources of contamination: terrestrial radio
signals from human-generated radio frequency interference (RFI) and
the Earth's ionosphere, and extremely bright astrophysical
synchrotron foregrounds from our own galaxy. The former can be addressed with suitable site
locations and benefits from RFI mitigation and ionospheric characterization work from current low frequency
instruments. We can address the latter by using the inherent spectral
smoothness of the foregrounds to separate them from the cosmological
signal. However, this places stringent requirements on frequency-dependent instrument calibration, and foreground removal becomes a key
design driver for instrument characterization, stability, and
uniformity. 

Next-generation power spectrum measurements of 21\,cm emission require larger arrays that must be capable of foreground removal and be sensitive enough to detect the faint cosmological signal.  They require a high degree of element redundancy: uniformly-spaced identical feeds, allowing
fast-Fourier transform (FFT) beamforming for data correlation and
compression, operating across a broad redshift range, 
and utilizing real-time gain calibration. As noted below, storing the full correlation matrix is not practical, but beamforming this number of detectors as a method of data compression is possible with present-day computation resources, although it requires real-time calibration that has not yet been demonstrated with current instruments. This input from current experiments is critical to assess the trade-offs between raw sensitivity and ease of calibration. Achieving
foreground removal requirements with a sensible analysis strategy can only occur with a concerted R\&D effort along
three directed paths, described in more detail throughout this section: 

\begin{itemize}
\item \textit{Technological:} The primary technological development paths to
  build and calibrate future instruments include \textbf{improved
    signal processing and digital conversion electronics};
  \textbf{optimized RF analog chain design with an emphasis on uniformity}; and \textbf{gain
    stabilization and beam characterization instrumentation}.

\item \textit{Analysis:} The primary analysis path is to build on the foreground removal and RFI mitigation techniques from current generation experiments and develop FFT beamforming compression and associated instrument design specifications to enable analysis at an achievable computation scale.

\item \textit{Simulations:} The primary simulation path is to build
  synthetic data for Development and Deployment, Validation and
  Verification, and Uncertainty Quantification. This must include full
  instrument characteristics to optimize instrument design and fully
  explore cosmological parameter constraints, particularly for analysis
  involving cross-correlations and other survey data. The minimum
  required inputs to form a sky map for this process are mock catalogs
  with galactic foregrounds and point sources. By the time this
  project becomes reality, our understanding of the low-frequency sky will
  be considerably improved from the current generation of EoR and post-EoR experiments.
\end{itemize}

In Section~\ref{subsec:technical_challenges} we review the outstanding design requirements for $21\,$cm cosmological mapping, heavily informed by the experience of the current generation of experiments. In Section~\ref{subsec:enabling_technologies} we summarize the main technological R\&D areas to address these, and then describe specific technology advances in more detail. In Sections ~\ref{subsec:data_analysis} and~\ref{subsec:simulation_challenge} describe the analysis and simulations challenges, respectively. Finally, in section~\ref{subsec:DOE_capabilities} we relate the technical needs of a $21\,$cm experiment to historical DOE strengths and capabilities, as well as pointing out opportunities for growth.

\subsection{Design Drivers and Requirements}
\label{subsec:technical_challenges}

\begin{figure}
  \centering
  \includegraphics[width=1.0\linewidth]{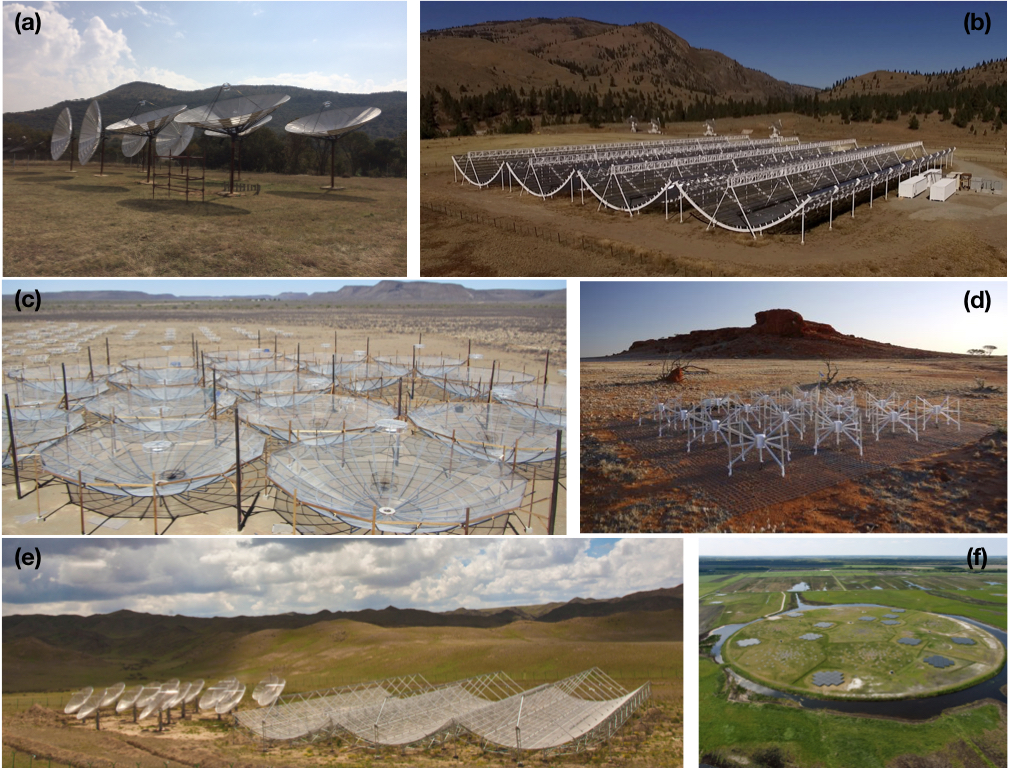}
  \caption{A sample of relevant $21\,$cm interferometric experiments currently fielded: \textbf{(a)} 8-element HIRAX prototype array operating at 400--800\,MHZ \cite{2016SPIE.9906E..5XN}; \textbf{(b)} CHIME experiment operating at 400--800\,MHz \cite{2014SPIE.9145E..22B}; \textbf{(c)} HERA \cite{2017PASP..129d5001D} operating at 50--250 \,MHz with PAPER \cite{pober_et_al2013b} in the background; \textbf{(d)} MWA operating at 80--300\,MHz \cite{2012rsri.confE..36T}; \textbf{(e)} Tianlai \cite{2012IJMPS..12..256C} operating at 700--800\,MHz; \textbf{(f)} LOFAR \cite{2013A&A...556A...2V} operating at 10--230\,MHz. EDGES \cite{2017ApJ...835...49M} is not included because it is targeting a global signal. LEDA is relevant but not pictured ~\cite{6328797}.}
  \label{fig:21cmexpts}
\end{figure}

There is now $\sim$ 1 decade of experience using radio telescopes to measure cosmological neutral hydrogen in `intensity mapping' mode \cite{2010Natur.466..463C,2013ApJ...763L..20M,2013MNRAS.434L..46S, 2017ApJ...847...64M}. Experience from current generations of experiments already taking data (HIRAX, CHIME, LOFAR, PAPER, HERA, MWA, and Tianlai among others -- see
Figure~\ref{fig:21cmexpts}) has shown that the most challenging
requirements come from tackling bright astrophysical
foregrounds. Astrophysical foregrounds, primarily synchrotron emission from the galaxy and unresolved point sources, have much higher
intensity than the cosmological signal of interest. Sobering experience from 21\,cm surveys over the past decade \cite{aguirre2019roadmap,CHIMEresults2022} shows that the current results are still limited by their ability to remove foregrounds. These foregrounds have a smooth spectral shape and hence can in principle be distinguished from the $21\,$cm emission from large scale structure~\cite{2017PASA...34...33P,pober_et_al2013b,seo_and_hirata2016,2012MNRAS.419.3491L}. However, any frequency dependence in the instrument response, for example from the instrument beam or gain fluctuations, can complicate our ability to differentiate between the smooth foreground and the essentially Gaussian cosmological signal~\cite{Shaw:2014vy,Shaw:2013tb}. Removing these foregrounds drives design choices including element uniformity, array redundancy, assessment of instrument stability and stabilization methods; provides opportunities for new calibration techniques in both beam and gain measurements; and requires analysis and simulations to fold in calibration measurements and assess their impact on cosmological parameter estimation. \textbf{ Below we outline the design drivers to overcome foreground contamination, which will require dedicated efforts across all of instrument design, data analysis, and simulation.}

\newcommand{\maybebullet}{}

\medskip
\noindent{\bf \maybebullet Instrument Calibration.}
Work in $21\,$cm calibration focuses on instrument gain and beam
measurement for the goal of removing astrophysical foreground
power. Simulations for CHIME have provided a scale to the problem: the
instrument response on the sky (`beam') must be understood to 0.1\%,
and the time-dependent response of the instrument (`gain') must be
calibrated to 1\% ~\cite{Shaw:2013tb,Shaw:2014vy}.  Current
instruments rely primarily on sky signals for both types of calibration, however
this has not yet been demonstrated to adequately remove foregrounds with these instruments. Throughout this chapter we outline design choices to meet uniformity and stability specifications that must be carefully integrated into the instrument design, verified during testing and deployment, as well as develop or advance new methods of calibration for this removal.

\medskip
\noindent{\bf \maybebullet FFT beamforming requiring real time calibration and array redundancy.}
The correlation cost and data rate from future large arrays 
such as PUMA will
require implementing an FFT beamforming correlator. Such correlators
use FFT-based sampling of the interferometric
geometry~\cite{Tegmark2009} to reduce the computational
correlation cost from order $N^{2}$ to $N\log N$ and output data
volume from $N^{2}$ to order $N$. Taking advantage of this technique
requires that all elements of the array be redundant (that their beams
are similar), placing stringent requirements on element uniformity. In
addition, this correlation will be performed in real time, and so this
requires that we employ real-time calibration to account for
instrumental changes (or that the instrument remains extremely
stable). This technique will be attempted on current generation
telescopes, and we expect work on those experiments will inform
requirements and algorithms for future 
instruments such as PUMA.

\medskip
\noindent{\bf \maybebullet Environmental considerations.} In addition to astrophysical foregrounds, two sources of terrestrial signal contaminants must be eliminated or otherwise mitigated: human generated radio-frequency interference (RFI) and Faraday rotation in the ionosphere.

Radio bands within the $21\,$cm redshift range $0.1<z<6$ are popular
as communications frequencies. This forms a bright RFI signal at
discrete frequencies within our measurement band. RFI can be reduced
or eliminated by a suitable choice of radio-quiet
observation site such as the middle of South Africa or western
Australia~\cite{2015PASA...32....8O}, which are remote areas with
limited communications in countries with suitable infrastructure. Even
if RFI must be removed, various experiments operating in locations
with high degrees of interference, notably LOFAR (located in the
Netherlands), have built impressive RFI removal techniques
\cite{2012A&A...539A..95O} that future 
experiments such as PUMA can draw from.

The ionosphere is a plasma and acts in concert with the Earth's magnetic field to rotate the polarization vector of incoming light. The rotation is proportional to $\lambda^{2}$ as well as the number of free electrons present in the ionosphere, which vary across all time scales. While the cosmological signal is unpolarized, most foreground emission from the galaxy is polarized, and so this adds a time variable component to the foreground characterization and removal. The $\lambda^{2}$ dependence means it is not expected to impact the shorter wavelengths (frequencies above 500\,MHz, $\sim z<2$), but it will impact longer wavelengths relevant to higher redshift experiments such as PUMA. 
Work towards measuring and removing this rotation using accurate maps of Earth's magnetic field and GPS data to infer free electron content is ongoing for experiments at long wavelengths. Because signal propagation through the ionosphere is critical for satellite telecommunications, it is well modelled and current low frequency radio telescopes are working to remove signal variability from the ionosphere~\cite{2015PASA...32...29A}.

\medskip

\noindent{\bf \maybebullet Required Sensitivity.} Instrument noise stems from a combination of intrinsic amplifier noise (noise temperatures for state-of-the-art radio telescopes range from 25\,K cryogenic to 100\,K uncooled) and sky brightness temperature (which span between
10\,K - 1000\,K depending on pointing and frequency). Because synchrotron emission increases at lower frequencies, at high redshifts
(above $z\sim$ 3) the system noise is dominated by the sky and no
longer by the amplifier, thus improved noise must be achieved by
fielding more antennas rather than better performing front-end amplifiers. In the absence of systematic effects, detecting the $21\,$cm signal requires fielding instruments including thousands of receivers to achieve mapping down to the mean brightness
temperature of the cosmological $21\,$cm signal of $\sim$0.1 - 1\,mK in the redshift range $0.1 < z < 6$ within a few years.

\medskip
\noindent{\bf \maybebullet Computing Scale.}
Radio astronomy has always been at the forefront of `big data' in
astronomy. Current generation $21\,$cm instruments produce $\gtrsim \SI{100}\,$TB of
data per day without any compression, natively generating an amount of data $\propto N^{2}$ where $N$ is the number of elements (currently $N\sim 10^{3}$),
representing challenges in data reduction, transfer, storage, analysis, distribution, and simulation. Compression by a factor of $\sim N$ is achievable by exploiting redundancy within the interferometer, but requires the use of real-time, in-situ calibration and places strong constraints on the uniformity of the optics between elements. To aid in data transport, analysis, and data quality assessment, data can be compressed further by co-adding maps.
This reduces the data size but increases pressure on real-time instrument calibration. In addition, to enable transient science we will need fast triggers (e.g., for detecting fast radio bursts), already deployed at current generation instruments.

\subsection{Technologies Enabling Science}
\label{subsec:enabling_technologies}

Understanding the instrument requirements illustrated above allows us
to identify dedicated, targeted, and coordinated research and  development areas that
will enable future $21\,$cm 
experiments such as PUMA to reach the science goals presented throughout this document. We propose a multi-pronged development effort: early digitization for improved stability and uniformity, optimizing the analog radio receiver elements, and new methods in beam and gain calibration.

\subsubsection{Early digitization and signal processing}
\label{subsubsec:Electronics}

\medskip

Most generally {\it gain} refers to the scaling between the incoming signal and the digitized signal, typically from the analog system (feed antennas, amplifiers, cables) and digitizer. Analog components are subject to gain
variation, typically due to temperature changes, as the signal travels
from the focus of the dish to the later digitization and correlation
stages. As noted above, gain variation is one of the limiting factors
in removal of astrophysical foreground power. One avenue of
development is to digitize directly at the focus of the dish because
signal information is ``vulnerable'' at all points along the analog
stages, so the imperative is to digitize as early as possible, after
which the signal is (nearly) ``invulnerable''. The resulting digital
signal has more resiliency against time-variable changes in the signal
chain (while some of these are simply moved from analog signal
transfer into the clock distribution, the latter is inherently
narrow-band), offers the possibility of more flexibility in
calibration injection signal algorithms to make gain solutions more
robust, and allows us to use commodity or other well-established
protocols developed for timing and data transfer. However, this comes
at the expense of overcoming the RFI from the digitization in the
field, potentially increased cost, and will require all amplification
to occur at the focus and thus we may find we need carefully designed amplifiers and thermal regulation
at the focus as well.

Several technology developments make receiver electronics with integrated digitizers (early digitization) a promising technology for $21\,$cm projects. Critical components that are now available commercially include:
\begin{itemize}
\item
Room temperature amplifiers with noise temperatures below sky brightness requirements from 100\,MHz to 1.2\,GHz.

\item
Low cost digitizers operating in the gigahertz regime with up to 14-bit resolution are readily available. This allows a trade-off: high bandwidth
direct digitization provides the ability to oversample and design high performance digital band selection filters and high order frequency equalizers, but analog conditioning is simpler to implement and model. The final
design will be decided by cost trade-offs while still meeting
stability requirements for foreground removal.
\item
Low cost programmable logic devices capable of interfacing with a high-speed ADC, providing digital filtering to the frequency range of interest, and interfacing to high speed networks.
\item
Similarly, the availability of integrated RF / ADC / FPGA devices in the near future may provide a path to very compact high-performance receivers.
\end{itemize}

By digitizing at the focus we broaden the possibilities for instrument calibration, bandwidth, and signal processing, however there are a few additional considerations:

\begin{itemize}
\item As noted, one of the technical challenges for $21\,$cm telescopes is the need for $<$1\% gain stability over at least 24 hours.  The primary culprits of gain variation with temperature come from the amplifiers and any analog transmission (either coaxial cable loss or radio-frequency-over-fiber). By digitizing at the focus, the analog transmission is unnecessary and then any variation will be dominated purely by the amplifiers. The resulting temperature variation can be either mitigated by use of thermal regulation of the circuitry at each dish focus or removed by injecting a calibration signal, or both. Because noise diodes have a gain stability of $2\times 10^{-3}/^\circ$C, achieving the required gain stability still requires thermal regulation of $\sim ^\circ$C. Amplifiers have roughly similar thermal regulation requirements, however they are more difficult to decouple from the environment because they are either connected or embedded in the antenna. Thus, development should be placed towards building calibration sources, digital or otherwise, to enable gain stabilization.

\item We must isolate the sensitive RF input with signals in the -100\,dBm range from the high power digital outputs from the ADC which typically operate near 0\,dBm. In addition, RF radiation from the digital processing system must be shielded from the input and from any other antennas.

\item The raw data rate from the digitizer is large, a few
  $\times$10\,Gbit/second. This can be substantially reduced depending
  on the oversampling level, with digital filtering in the FPGA that
  receives the digitizer data, followed by transmitting only the
  bandwidth containing useful physics data. For some correlator
  architectures it may also be useful to transmit data separated by
  frequency band to an array of correlation processors. The system can
  trade off oversampling at a few gigahertz and digitally filtering
  down to the band of interest for a more complex analog system.  In
  theory a digital filter can do significantly better than an analog
  filter in terms of stability and out of band rejection, and may
  become more cost effective on the time scale of future 
  instruments such as PUMA.

\item Digitizing at the focus simplifies and stabilizes the signal handling down-stream, however clock jitter between components would decrease the correlation efficiency and thus impact signal stability and recovery. To mitigate this, the clocks driving the ADC would need to be synchronized to sub-ps precision, which is several orders of magnitude below the sampling interval. Conversely, if such synchronization is achieved, the entire system gains orders of magnitude in stability -- it does not matter if the data packets arrive somewhat later due to thermal expansion in the cables as long as they are correctly time-stamped. Luckily, synchronized timing is critical to many experiments, and photonic phase synchronization systems (for example baselined for SKA-MID~\cite{1805.11455}) achieve $\sim 50\,$fs stability over minute timescales with direct application in the radio telescope domain, while more sophisticated versions claim $10^{-17}$ stability over hour timescales \cite{Wang:19}.  

\end{itemize}

\subsubsection{Analog design}
\label{subsubsec:optics}

\medskip
\noindent
For power spectrum measurements of the faint 21\,cm signal, the instrument is driven towards a many-element wide-band close-packed interferometer with a high degree of baseline redundancy and uniformity between dishes to boost signal to noise on the faint cosmological signal. There are a few challenges in this design, in particular:

\medskip
\noindent
\textbf{Wide and uniform radio bandwidth -- } Interferometric experiments typically have a \textit{minimum} bandwidth of 2:1, and can range as high as 5:1.  This drives the feed antennas to be large, which can present significant obstruction (and hence lower collecting area) of the reflector if the feed is positioned at the focus. To decrease cross-coupling between dishes, arrays may also choose low-$f$-number, fast telescopes, which allow the primary focus to ``hide'' below the edge of the parabolic surface, thus minimizing the coupling between elements. This mildly exacerbates the obstruction from the feed. Finally, the differential response to linear polarization components needs to be well controlled. The foregrounds to the 21\,cm signal (described in more detail below) are polarized via Faraday rotation in the inter-stellar medium, which is also frequency dependent. To measure and remove polarized foregrounds, the polarized response of the telescope must be well understood across the entire band, and ideally the spectral response would be designed to be smooth.

\begin{figure}
  \centering
  \includegraphics[width=0.5\linewidth]{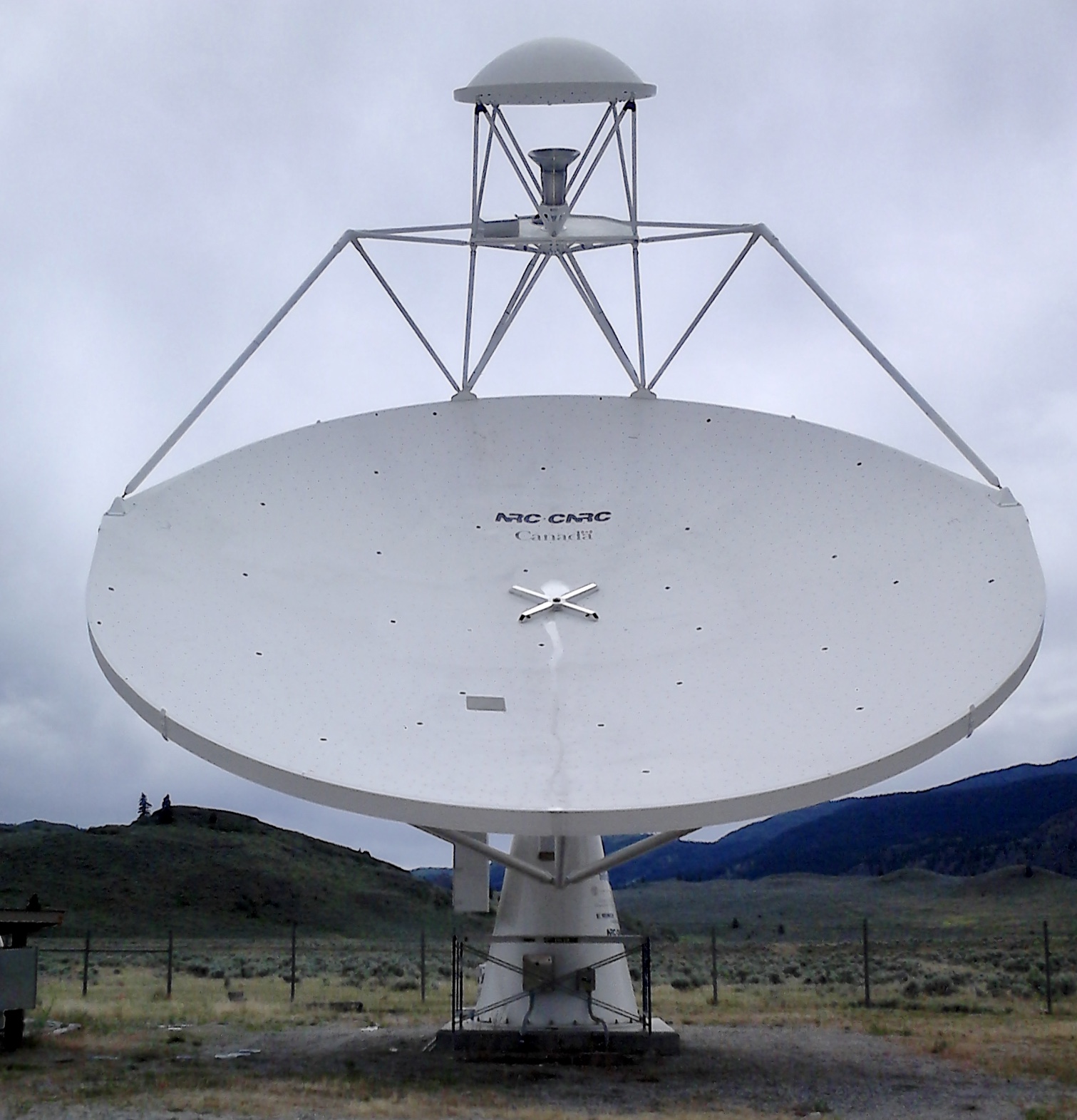}
  \caption{Prototype SKA fiberglass dish, located in Canada at the Dominion Radio Astrophysical Observatory}
  \label{fig:fiberglass}
\end{figure}

\medskip
\noindent
\textbf{Uniform reflector elements -- }
Uniformity and stability of individual antennas is of crucial importance for calibration and efficient data processing and compression. Significant progress in using carbon fiber composite dishes for repeatable dish shapes has been made for the SKA and cheaper fiberglass alternatives are being prototyped for the future 21\,cm arrays HIRAX and CHORD (such as using a fiberglass-based design~\cite{7303193}, see Figure~\ref{fig:fiberglass}). At a minimum, experience and testing from current and upcoming experiments will inform future arrays, and may also drive us towards developing alternatives (e.g.\ fiberglass reinforcement, hard foam, etc.) to meet specifications.

\medskip
\noindent
{\bf \maybebullet Front-end sensitivity and bandpass}.
When properly designed, the receiver noise temperature is dominated by loss in the analog feed as well as the noise in the first stage amplification. HIRAX has chosen to reduce the system noise by up to 30\% by fabricating the first stage amplifier directly in the antenna itself, reducing the transmission loss and taking full advantage of low-noise transistors available in these bands. In addition, current generations of $21\,$cm experiments~\cite{2017PASP..129d5001D} have found that their bandpass shape is a limitation of their foreground removal, and are actively working on new feed designs that have a more carefully shaped bandpass. One development path for the active circuitry in the HIRAX feed would be to add additional RF circuitry to flatten the bandpass to remove ripple and other features, allowing an easier path for foreground removal. This introduces more stringent oscillation conditions on all amplification stages to reduce the possibility of amplifier oscillation and we will learn more about the feasibility of this technique for mass production as additional prototypes are fabricated for HIRAX. \newline

\medskip
\noindent
{\bf \maybebullet Uniform interferometric elements for calibration and FFT correlation.}
Similar interferometric baselines should see the same sky signal and so differences between them can be used to assess relative instrument gains over time. This technique is known as `redundant baseline' calibration and has been developed as a method of meeting the gain stability requirements~\cite{2010MNRAS.408.1029L,2018arXiv180500953C,2018MNRAS.477.5670D,2014ASInC..13..393R,2016ApJ...826..181D,2017arXiv170101860S}. This requires both a decision to space the interferometer dishes the same distance apart, and also have highly uniform interferometric elements. Most $21\,$cm instruments have chosen their baseline spacing to use this technique, however have been limited by the fact that their interferometric elements are not identical enough to achieve precision calibration. To overcome this, we would investigate dish fabrication tolerances required for this calibration as well as how we might use new dish fabrication techniques (for example, fiberglass dishes with embedded mesh conductors, currently being prototyped for SKA and HIRAX, see Figure~\ref{fig:fiberglass}) to meet these needs.

In addition, the requirements that we use FFT or similar beamforming~\cite{2004NewA....9..417P,TegmarkZaldarriaga2010,2017arXiv171008591M} to compress that data forces stringent requirements on the uniformity of response, beam shape, mechanical construction and alignment, gain control, etc.~across what will ultimately be on the order of $\sim$65k detector copies. The requirements for this uniformity and how to achieve it will be part of the instrument design process.

\subsubsection{Instrument Calibration}
\label{subsubsec:beam_characterization}

\medskip
\noindent
{\bf \maybebullet
Gain Stability.} Each antenna has a characteristic response to an input sky signal, which varies with both time and frequency, known as the instrument gain. The frequency-dependent gain for each input must be known to $\sim$ 1\% on time scales between the integration period ($<5$s scales) and a few hours (depending on the frequency of on-sky radio calibrator sources)~\cite{Shaw:2014vy}. The two primary techniques for achieving this are to design an instrument which is inherently stable enough to meet this specification or to design a calibration plan which can ensure we meet this specification, or (ideally) both. CHIME~\cite{2014SPIE.9145E..22B,2014SPIE.9145E..4VN} is updating a classic radio noise-injection scheme which can be used to calibrate many signal chains at once. To implement such an active calibration technique for dishes will require development of stablized transmission algorithms and may be made easier with early digitization and development of calibration sources which may be independently fielded at the focus or flown on a quadcopter drone. We will also require passive models of gain and beam variation with temperature and dish pointing. This modeling is essentially standard for radio telescopes and precision modelling has been demonstrated with at least one instrument (CHIME).

\medskip
\noindent
{\bf Beam Characterization.} Each antenna also has a characteristic response on the sky, known as the instrument beam. Because this response (main beam and sidelobes, as well as polarization) is capable of mixing frequency dependence and sky location, it is expected to be the primary source of contamination from foreground emission into the signal band. As a result, preliminary simulations indicate we must know the beam to a precision of $\sim0.1\%$ \cite{Shaw:2014vy} across the entire band to enable adequate foreground removal. Precautions must be taken to avoid internal resonances or reflections that introduce non-smooth spectral structure into the instrument response. Current-generation experiments are developing the calibration strategy for transit interferometers. Their approach uses a variety of measurements: celestial sources that can be reliably measured at transit \cite{CHIMEresults2022}; measurements of the sun (with solar variability correction extrapolated from solar telescopes) \cite{CHIME_sun_2022}; pulsars (where their flashing allows removal of all confusion noise); and satellites (which are narrow band, and usually variable during a beam crossing)\cite{NebenOrbcomm2015}. Many $21\,$cm instruments are beginning to use quadcopter drones to map the beam shape (HERA\cite{Jacobs_drones}, SKA\cite{Virone_drones, Pupillo_drones}, LOFAR\cite{Chang_drones}) and while this technique seems promising to meet the needs for $21\,$cm cosmology it is unlikely we will be able to measure all of the beams from all of the dishes, and so this beam calibration requirement also forces a specification on uniformity in dish fabrication.

Current problems identified from a variety of groups include overcoming the dynamic range between the main beam and the lower sidelobes (sources bright enough to measure the sidelobes saturate the main lobe, reducing the ability to `stitch the measurements together' for accurate maps). One solution is to fly multiple times with a variety of attenuation levels, but this requires far greater flight times. In addition, contributions from the changing sky brightness will impact measurements at the 0.1\% level, for which flashing the source is a good approach but one that also requires additional flight and integration time. A variety of approaches are being considered, including digital calibration sources \cite{2022arXiv220111806B} which may also be used for gain stabilization if digitization at the focus is successful and fixed wing drones that are considerably faster and with longer flight duration.

\subsubsection{Data flow and processing}
\label{subsubsec:data_flow}


Computing requirements for future large interferometers come from both the correlation burden and the data reduction, transfer, storage, analysis, and synthetic data production. For the correlator computation, we will need to pursue development in computing approaches which
can improve the cost scaling both for equipment and power. Examples could
include using commodity-hosted FPGA's, using/developing dedicated
ASIC's \cite{7436238}, or GPUs to smoothly take advantage of the fast-paced hardware updates for correlator computation.

\subsection{Data Analysis}
\label{subsec:data_analysis}

\newcommand{\tcite}[1]{[{\color{green}#1}]}  
\newcommand{\rscomment}[1]{{\color{red} \textbf{[RS:  #1]}}}


\medskip

Releasing science deliverables for the community from a $21\,$cm experiment depends crucially on developing and deploying, including validation and verification, an analysis pipeline that can ingest vast quantities of data and transform it into well characterized frequency maps and power spectra. This is a computationally costly and varied exercise, but does not require continuous real time processing, and thus can be performed at an external high performance computing site.
We can divide the analysis up into three broad areas discussed below.

\medskip
\noindent {\bf Flagging, Calibration and Pre-processing at scale.} In this area
the data is processed to reduce the remaining systematics which may
effect our ability to access the cosmological signal.
Of particular importance is cleaning of any RFI by flagging times and frequency channels that have been contaminated. This is a well understood problem within radio astronomy \citep{2012A&A...539A..95O}, though the effects of residual RFI at the small level of the $21\,$cm signal is only starting to be addressed \citep{2018MNRAS.479.2024H}.
Though much of our calibration must be done in real-time (see \ref{subsec:technical_challenges}) to enable FFT correlation, there are still degrees of freedom that must be corrected, particularly degeneracies that may not have been fully fixed (including but not limited to an overall gain scale for the entire observation, \citep{2010MNRAS.408.1029L}), and calibration of the bandpass (the array-wide frequency dependent gain). Again these are problems that are well understood within radio astronomy.\\[-0.8em]

\noindent {\bf Astrophysical Foreground Removal.} Along
with the sensitivity requirements for measuring a faint signal, the
key analysis problem for $21\,$cm intensity mapping is the need to
remove contaminants that are many orders of magnitude larger than the
cosmological signal. Though foreground cleaning is a common problem
across cosmology, the required dynamic range is unique to $21\,$cm intensity
mapping.

In principle the foregrounds can be separated from the signal using their smooth frequency dependence \citep{2010ApJ...721..164S}. However, even an ideal instrument couples anisotropy in the astrophysical foregrounds into spectral structure with an amplitude generally significantly larger than the cosmological signal. This extremely challenging problem is called mode-mixing and is exacerbated by instrumental systematics such as gain variations and optical imperfections which must be minimised (see the discussion in Section~\ref{subsec:technical_challenges}). There exist in the literature many proposed techniques to separate the cosmological signal from the foregrounds, but these have only demonstrated success in simulations.

Foreground mitigation falls broadly into two classes: foreground
avoidance and foreground cleaning. Foreground avoidance is the
simplest of these two approaches, relying on the fact that
contamination produced by a typical interferometer configuration is
strongest in certain regions of $k$-space. Producing cosmological results only using the cleanest modes is a simple and effective technique. This technique, however, becomes deeply unsatisfactory at low frequencies, particularly in the Dark Ages. Here galactic synchrotron and extragalactic point source radiation quickly becomes very bright, typically hundreds of Kelvin at \SI{100}{\mega\hertz}, even at high galactic latitudes. At the same time the window of clean modes dramatically narrows due to the relative scaling of the angular diameter distance and Hubble parameter with redshift \cite{2015MNRAS.447.1705P}. Combined, this means that at a given threshold for contamination we exclude increasingly large regions of $k$-space at high redshifts, significantly degrading any cosmological result.

Foreground cleaning instead of (or in conjunction with) foreground avoidance then becomes an attractive option. A general feature of foreground cleaning methods is that they rely on detailed knowledge of the instrument response to predict and subtract the actual foreground signal. For instance, given perfect knowledge of the complex beam of each individual antenna, a tomographic map of the sky can be effectively deconvolved to remove the spectral structure induced by the instrument's beam. The residual contamination is set by both the amplitude of the raw contamination and the accuracy with which the beam has been measured. This is similar in spirit to the residual temperature-to-polarization leakage produced by mismatched beams of orthogonal polarizations in CMB $B$-mode searches, which can be accurately predicted and removed given beam measurements despite the fact that the CMB temperature anisotropy ``foreground'' is orders of magnitude larger than the $B$-mode signal.\\[-0.8em]

\medskip
\noindent {\bf Cosmological Processing.} Having cleaned the foregrounds out of the data we then need to process it to quantities useful for cosmology such as power spectra and sky maps. Though this has been done within the CMB and LSS communities for many years, the fact that we are dealing with radio interferometric data after foreground cleaning brings unique challenges. The source of these is that the measured data is abstract: it is a complex, spatially and spectrally non-local measurement of the sky. This adds significant complexity in generating maps and power spectra from the data, but also tracking which modes have been measured (and which are missing) to allow us to accurately measure uncertainties. Regardless, we expect to be able to significantly draw on the conceptual frameworks used for cosmological data analysis to be able to tackle these problems \cite{Shaw:2014vy,2014PhRvD..90b3018L,2004ApJ...609....1J}.\\[-0.8em]

Although we can create a broad outline of how the analysis pipeline,
and we are able to draw on many mature and well understood techniques,
there are several areas that will require research investment to
ensure the success of a large scale $21\,$cm intensity mapping survey.

\medskip
\noindent {\bf Scaling.} While we can draw on existing techniques for all stages of the analysis, a significant challenge is scaling these to be able to work with the vast increase in data that we will generate in an energy-constrained/post-Moore's computing landscape. This will require optimizations in algorithms and implementations to reduce the computational cost of the processing, and ensuring that the techniques can scale in parallel to run on leading edge supercomputers.

\medskip
\noindent {\bf Systematic Robustness.} Both astrophysical uncertainties (such as the exact nature of foregrounds) and instrumental uncertainties (such as calibration and beam optics) cause foreground contamination. Developing more robust cleaning techniques will reduce systematic biases, but potentially allow us to reduce the instrumental tolerances leading to cost savings.

\medskip
\noindent {\bf Improving signal recovery.} Significant numbers of modes are lost to foregrounds, which reduces our constraining ability generally, but particularly affects science that needs access to the largest scales. Improved foreground removal that reduces the effect of the wedge could improve this, as would methods like tidal reconstruction \cite{Zhu:2015zlh,Zhu:2016esh,Foreman:2018gnv}, but these techniques need substantial development. Similarly, traditional reconstruction techniques \cite{2007ApJ...664..675E,2009PhRvD..79f3523P} that recover non-linear modes need work adapting them for the peculiarities of $21\,$cm intensity mapping.

\subsection{Simulation Needs and Challenges}
\label{subsec:simulation_challenge}

The challenges facing $21\,$cm surveys are significant but, at least to $z=6$, well understood. However, our ability to tackle them requires a sophisticated approach to overcome them both through instrumental design and offline analysis. It is therefore essential to use simulations to close a feedback loop that allows us to predict, and thus refine, the effectiveness of a design and analysis strategy.

Producing realistic simulations of data from any instrument configuration and propagating these to final cosmological results is a conceptually straightforward prospect:
\begin{enumerate}
  \item Produce a suite of full-sky maps of the ``true'' sky, with one map per frequency and at each frequency bin observed by the instrument. There are a variety of approaches to form full-sky maps of the signal and foreground, and full exploration of the data should include common sky models to include other observables (e.g. galaxy surveys) for form estimates of cross-correlations. 
  \item ``Observe'' these maps with a simulation pipeline that contains sufficient realism to capture any and all non-idealities that might produce contamination in the data.
  \item Feed these mock observations into the data analysis pipeline discussed in the previous section, and the same pipeline that would be used on real data, and produce reduced data and cosmological analyses.
\end{enumerate}

\begin{figure}
  \centering
  \includegraphics[width=0.49\linewidth]{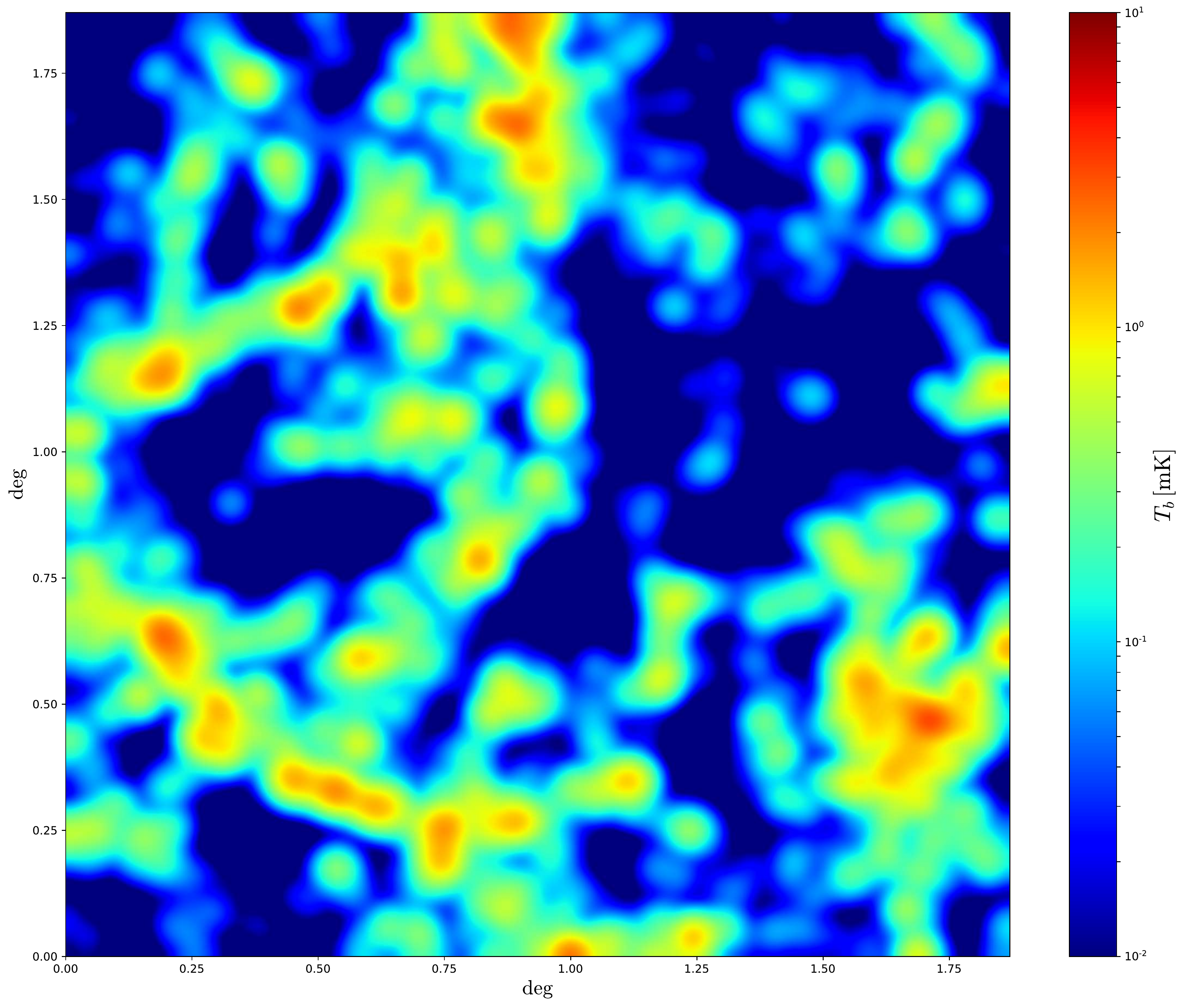}
  \includegraphics[width=0.49\linewidth]{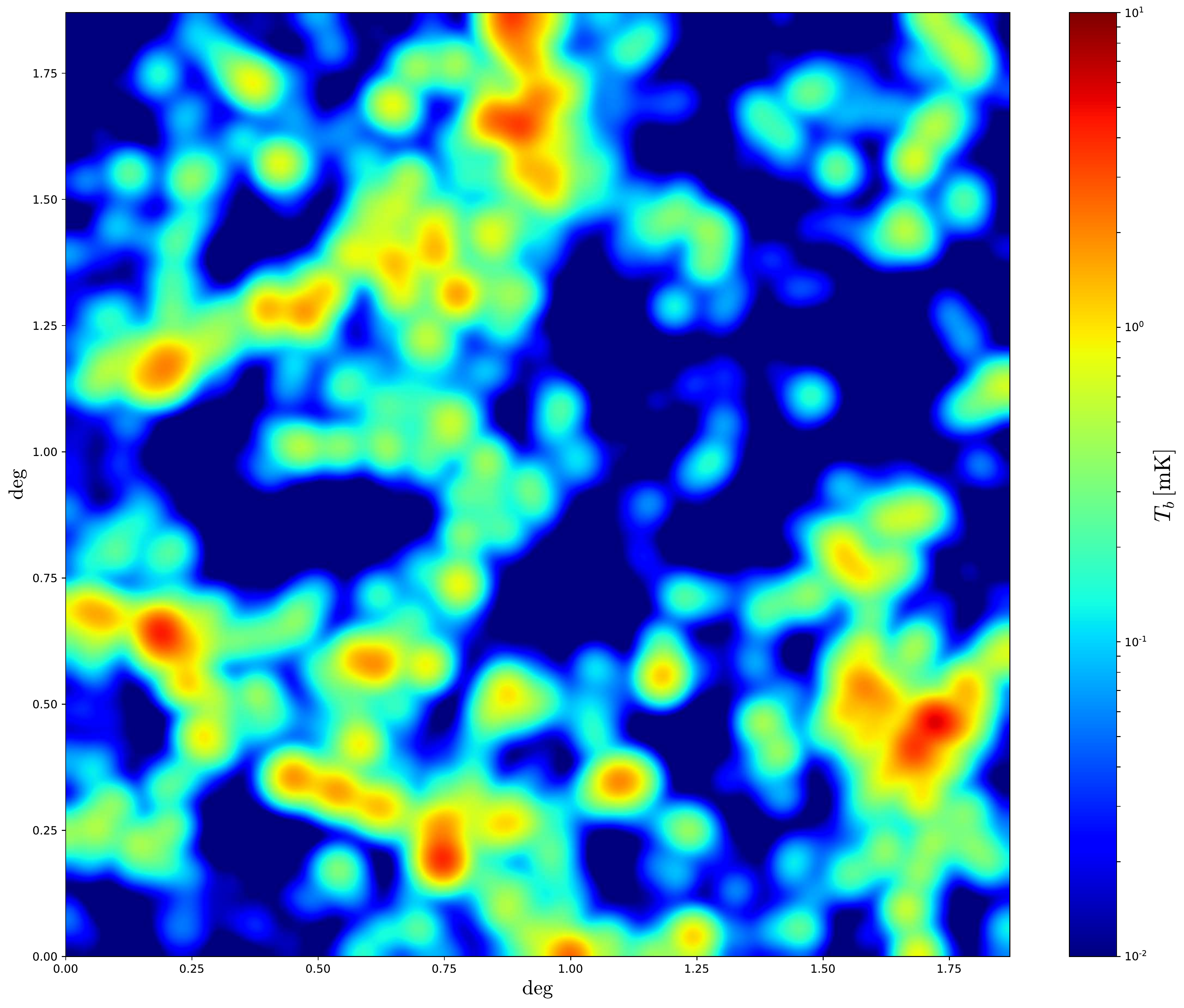}\\
  \caption{21cm maps at a frequency of 710 MHz over a channel width of 1 MHz with an angular resolution of 1.5' over an area of $\simeq4~{\rm deg}^2$. The map on the left has been created from the state-of-the-art magneto-hydrodynamic simulation IllustrisTNG with a computational cost of $\simeq$18 million cpu hours. The map on the right panel has been generated by assigning HI to dark matter halos of an N-body simulation using the a simplification of the ingredients outlined in \citep{2018ApJ...866..135V}. The computational cost of the N-body simulation is much lower than that of the full hydrodynamical simulation, and allow us to model the HI field in a very precise and robust manner.}
  \label{fig:21cm_sim}
\end{figure}

For verification of foreground removal effectiveness Gaussian or pseudo-Gaussian $21\,$cm simulations are largely sufficient \cite{2014MNRAS.444.3183A,Shaw:2014vy}. However, for targeting sensitivity to specific effects (e.g. non-Gaussian initial conditions), or in cross-correlation with other probes, more accurate simulations constructed from mock-catalogues will be required. This allows us to produce correctly correlated maps for additional tracers (e.g. LSST photometric galaxies), and also for radio point source contribution to the foregrounds.

Though the relation between HI density and total matter density involves complex environment dependent processes, simulating it can be done efficiently. Recent work has shown that one can take advantage of the fact that neutral hydrogen in the post-reionization era resides almost entirely inside dark matter halos~\citep{2018ApJ...866..135V}. It is possible to combine the presently available HI data to constrain an analytical halo model framework for the abundance and small-scale clustering of HI systems \citep{Padmanabhan2017a, Padmanabhan2017b}. Thus, one can calibrate the relation between dark matter halos and HI using hydrodynamic simulations and create 21cm maps via less expensive methods such as N-body or fast numerical simulations like Pinocchio \cite{Pinocchio}, ALPT \cite{ALPT}, HaloGen \cite{HaloGen}, EZMock \cite{EZmock}, PATCHY \cite{Patchy}, COLA \cite{COLA}, QuickPM \cite{QuickPM}, FastPM \cite{FastPM}, or Peak Patch~\cite{stein-peak-patch}.  It may even be possible to adopt a simple perturbation-theory-inspired approach \cite{Modi:2019hnu}, which would allow very large volumes to be simulated very efficiently.

As the dominant foreground contribution, simulating the galactic synchrotron must be done with care to ensure that the simulations are not artificially easy to clean. A simple approximation can be produced by proceeding from a full sky map at a radio frequency (typically the Haslam radio survey map at 408~MHz) and scaling this map to different frequencies based on the known spectral index of galactic synchrotron radiation. However this is not sufficient at the dynamic range between the foregrounds and the $21\,$cm signal and we must be careful to include: spectral variations about a pure power law; small scale angular fluctuations not captured in existing surveys; and polarization, including the effects of emission at a wide range of Faraday depths which generates significant spectral structure in the polarized emission \cite{Shaw:2014vy}. More sophisticated galactic models, for example from MHD simulations, could also be developed and used here.

Regarding (2), a realistic instrument simulation pipeline would take
the maps discussed and convolve them with the complex beam for each
antenna in the interferometer. This can be done by direct convolution
utilising the fact that for a transit telescope it is sufficient to
generate a single day of data. However for wide-field transit
interferometers this can be more efficiently performed in harmonic
space using the $m$-mode formalism ($O(N\log{N})$ instead of
$O(N^2)$). Some of the required code would be similar and could in
principle built upon similar codes used in the CMB science, such as
the TOAST
package\footnote{\url{http://hpc4cmb.github.io/toast/intro.html}}
using fast numerical techniques for beam convolution
\cite{2010ApJS..190..267P}.

For these simulations we need to generate realistic simulations of the telescope beams. Electromagnetic simulation codes such as CST, GRASP and HFSS can be used for this, but achieving the accuracy required is challenging computationally \cite{2017JAI.....650003C,2017arXiv170808521D,2016ApJ...831..196E, 2021SPIE11445E..5OS}. An alternate approach is to generate synthetic beams with sufficient complexity to capture the challenges posed by real beam, these are computationally easier to produce, but must be informed by real measurements and electromagnetic simulations to ensure their realism, and may be aided by machine learning algorithms.

Capturing non-idealities in the analog system, particularly gain variations, is mostly straightforward as these can be applied directly to the ideal timestreams. Additionally we need to include time-dependent beam convolution (including position and brightness) for temporally varying sources such as solar, jovian and lunar emission as well as the effects of RFI at low levels~\cite{2018MNRAS.479.2024H}.

Including calibration uncertainties poses a particular challenge, because of the realtime calibration and compression of the instrument, simulating these effects requires either: generating data at the full uncompressed rate, applying gain variations, and then performing the calibration and compression processes; or the computationally easier alternative of generating models of the effective calibration uncertainties.

After the first two stages, mock observations are then fed to the proposed data analysis pipeline, and propagated through to final cosmological products, to assess analysis systematics, instrument design, real-time calibration, and data processing to determine whether the pipeline is sufficient to meet our science goals.

Though the simulation program is well defined, there are already many
open challenges discussed below.

\medskip \noindent {\bf Understanding the HI distribution.} To map the HI distribution to the cosmologically useful matter distribution requires cutting edge hydrodynamic simulations to capture the small halos that HI favours over a cosmologically interesting volume. This additionally improves the efficiency and accuracy with which we can produce mock skies.

\medskip \noindent {\bf Scale.} To assess and understand a proposed design we need to be able to produce large numbers of emulators that Monte-Carlo over the experimental uncertainties. The number, size and complexity of these simulations requires a large scale effort to plan, generate and manage them. 

\medskip \noindent {\bf Improving the Feedback Loop.} While a straightforward version of the simulation loop above can tell us whether a proposed design does or doesn't meet our needs, it does not tell us how to improve the design to ensure that it does. For a complex instrument with many design parameters is is essential to be able to guide this process by using simulations to infer the most relevant combinations of changes.

\subsection{Relation to DOE capabilities}
\label{subsec:DOE_capabilities}

 This chapter has enumerated the technology and analysis challenges for studying cosmic acceleration by mapping the large-scale structure of the Universe using $21\,$cm radiation. As with other large DOE-HEP experiments, it requires data from imperfect detectors to be turned into useful scientific output by application of multi-level calibration schemes that incorporate the as-built instrument characteristics and thorough end-to-end numerical simulation of the physics of the measurement process. DOE has a unique heritage in successfully constructing large experiments of this type, making it a particularly appropriate home for the development of a $21\,$cm intensity mapping experiment. Capabilities found in the DOE Laboratory complex in technical, computing, and management categories are discussed below.

\medskip \noindent \textbf{Technical capabilities.} DOE has long experience with RF systems for its hadron and electron accelerators. Hardware for manipulating RF modes to efficiently couple sources to accelerator waveguides and cavities has much in common with the matching optics used for radio telescope receivers. High channel-count, fast RF digitization and processing is also used extensively in control and beam diagnostics.
 Large accelerators such as LCLS-II can include over
a thousand channels of RF front ends and high-performance digitizers
connected to a distributed data network.  Although optimization of dynamic range, bandwidth, and noise characteristics differ from those needed for the $21\,$cm experiment, many commonalities between the designs remain.

Data acquisition systems at large HEP and photon science experiments generate
enormous volumes of digital data that must be transported over networks that
may comprise tens of thousands of high-speed links. Data transport, real-time
processing, and interface to commodity server farms requires DAQ developers to
have specialized expertise in the most modern microelectronics families
(ASICs, FPGAs, optical transceivers, etc.) and to be aware of rapidly
advancing trends that open opportunities for greater performance in future
projects. The Front End Link Exchange (FELIX) and global Feature Extractor
(gFEX) platforms being developed for the ATLAS experiment are examples of
state-of-the-art hardware coming out of the DOE labs; evolved versions of such
platforms can find very direct applications in real-time $21\,$cm signal
processing.

Table \ref{tab:datarates} shows a comparison of data rates in some current and future
experiments drawn from HEP, photon science, and radio astronomy. Data rates in a future 
 $21\,$cm experiment such as PUMA, 
 although challenging, are not out
of the range of some of the more ambitious projects shown.

\begin{table}
  \centering
  \renewcommand{\arraystretch}{1.4}
  \begin{tabular}{|l|l|l|l|l|}\hline
\multirow{2}{*}{Experiment}  & Data rate &  \multirow{2}{*}{~Year~~~} & \multirow{2}{*}{Note} & \multirow{2}{*}{Ref.} \\
& [GB/s] & & & \\ \hline
VLA & 0.3 & 2013 & Resident Shared Risk Observing mode & \cite{VLAdr}\\
ALMA & 1.8 & 2021 & Overall & \cite{ALMAdr}\\
LHC & 25 & 2018 & Average rate, all 4 experiments after triggering~~ & \cite{LHCdr}\\
LSST & 6.4 & 2022 & Peak rate & \cite{LSSTdr}\\
LCLS & 10 & 2009 & CXI instrument & \cite{Thayer2017}\\
LCLS-II & 320 & 2027 & High frame-rate scattering detector & \cite{LCLSIIdr}\\
XFEL & 13 & 2017 & 2D area detector & \cite{XFELdr}\\
SKA1 & 8,500 & 2022 & Overall & \cite{SKAIdr}\\
CHIME & 13,000 & 2017 & Input to F-engine & \cite{CHIMEdr}\\
\textbf{PUMA}~ & \textbf{655,000} & 2030 & \textbf{Input to F-engine} & \cite{2020arXiv200205072C} \\
\hline
  \end{tabular}
\caption{Rates for current and proposed data-intensive experiments, drawn from HEP, photon science, and radio astronomy. The data rates going into the F-engine for a future experiment such as PUMA are expected to be manageable within the time-frame of the experiment. \label{tab:datarates}
}
\end{table}

\medskip \noindent  \textbf{Computing capabilities.}
All stages of developing these experiments will require the
involvement of large computing facilities. The full system simulation as
well as actual data processing will require high-performance computing
and efficient storage, handling and processing of data volumes in the petabyte
range. This can be efficiently addressed through existing and planned
infrastructure facilities within the DOE laboratory complex that will
also drive new developments in network connectivity between DOE
sites. DOE runs NERSC, one of the world's largest high-performance
computing systems, ALCF and ORCF (limited access) and has put significant investment into exascale
computing across all centers. It also
hosts two CERN Tier-1 data centers.

In addition to challenges presented by the data volumes alone, there
are massive algorithmic challenges that can be efficiently addressed
using existing DOE structures present within Advanced Science Computer
Research (ASCR) and SciDAC. On the simulation side these includes
running large simulations of the Universe. On the data analysis sides,
the calibration problems and foreground removal problems can be recast
in terms of large-scale linear solvers, error analysis, kernel estimation,
machine learning, etc.  These problems will benefit from developments
in the current exascale initiative and work that has been done on
hybrid compute architectures that can be particularly efficient ith
large data rates.

\medskip \noindent \textbf{Management capabilities.}
A future $21\,$cm experiment such as PUMA will need to follow organizational models similar to
those that have evolved in DOE's other recent HEP programs. These may include
coordination with other agencies and/or international partners, setting up
scientific collaborations and a formal structure responsible for executing the
project plan, and arranging for appropriate levels of oversight. During the
construction phase, test systems for quality assurance and metrology will be
essential for mass-produced components to meet performance requirements.
Predecessor projects such as US-ATLAS/CMS silicon detector modules, LSST focal
plane raft towers, and CMB-Stage 4 detectors and readout will provide
useful models and lessons. Finally, DOE has experience in organizing
collaboration-wide scientific activities to generate high-fidelity simulations
of system performance. The LSST Dark Energy Science Collaboration's Data
Challenges are a recent example. As stated earlier, it will be absolutely
essential to perform end-to-end simulations for future experiments such as PUMA.

\medskip \noindent \textbf{Current DOE laboratory efforts.} There are currently
several small path-finder efforts at various labs not directly funded by DOE HEP.

Several DOE laboratory groups have been advocating for Packed Ultra-wideband Mapping Array (PUMA), a proposal for the next generation. See discussion in Section \ref{subsec:PUMA}.

At BNL, a small test-bed experiment, BMX, has been set-up operating at
1.1-1.5GHz \cite{oconnor20}. It has been taking data since Fall 2017 in single dish
mode and was upgraded to a 4 dish interferometer in the Summer
2019.  The results are promising despite the experiment being situated
at the lab site, which is an extremely poor location in terms of
RFI. Early science results include characterization of out-of-band
emission from global navigation satellite services that will act as a
potential systematic for low-redshift $21\,$cm experiments. As a
test-bed, the system will be used to test various approaches towards
beam and gain calibration and to gather on-sky data from early
digitization prototypes. It will thus continue to provide a convenient
bridge between laboratory testing and a test deployment on a real
radio telescope which often involves significant travel costs and
limited time allocation.

The Fermilab Theoretical Astrophysics Group has been closely involved
with $21\,$cm intensity mapping for the past decade.  Early work
included forecasting and technical design studies for $21\,$cm arrays
\cite{2010ApJ...721..164S} and development of analysis techniques
\cite{Shaw:2014vy,Shaw:2013tb}.  Currently, with NSF support, the
group hosts the Tianlai Analysis Center (TAC), which analyzes data for
the Tianlai instrument in China.  The current, ``Pathfinder" version of
Tianlai includes an array of 16, 6-meter diameter dishes and 3, 15 m x
40 m cylinder telescopes operating in the 650-1420 MHz range and acts
as a useful test-bed instrument for future efforts \cite{Das2018}.
Near-term goals include determining the optimal design of future
arrays (cylinders, dishes or both), and detecting HI at low and high
redshift ($z \sim$ 0.15 and 1.0).  The effort includes data storage,
calibration, RFI removal, data quality assessment, mapmaking, power
spectrum analysis, and development and testing of the Tianlai analysis
pipeline.  These tasks are partly enabled by the substantial computing
resources at Fermilab's Scientific Computing Division.

\section{Current and Proposed Projects}
\label{sec:facilities}

\subsection{LuSEE-Night}
LuSEE-Night (Lunar Surface Electromagnetics Experiment at Night) is a collaboration between DOE and NASA to install a test-bed radio receiver on the far side of the Moon. It will be launched on the Commercial Lunar Payload Services flight CS-3 in early 2025. It will consist of 4 monopole antennas that actuate in altitude and azimuth. The instrument will measure full signal correlations from the available antennas. 

The lander and all the other experiments will become electrically inactive during the first lunar dawn and LuSEE-Night will employ strong mitigation techniques to prevent self-contamination with radio frequency interference. The current plans include a far-field calibrator on the satellite that will deliver a coded signal from orbit and enable precise characterization of instrumental response, including that of the lunar regolith. 

LuSEE Night is expected to acquire data for at least 12 lunar nights. At the end of the mission it will provide the most exquisite measurements of the low-frequency radio sky below 50~MHz to date and demonstrate the feasibility of Dark Ages cosmology from the far side of the Moon.  The aim of the experiment is to characterise the radio sky at frequencies 1-50MHz with percent level absolute calibration and a $10^{-3}$ relative calibration between frequency bands. LuSEE-Night should have sufficient sensitivity to exclude presence of a monopole signal at about Kelvin level, a 1-2 orders of magnitude above the expected signal and sufficient to put constraints on some models predicting non-standard properties of baryon therodynamics during the Dark Ages.

\subsection{Monopole EoR experiments}

The Experiment to Detect the Global EoR Signature (EDGES) is a 21cm instrument, located at the Murchison Radio-astronomy Observatory in Western Australia ($26^{\circ}41^{'}50^{"}$S, $116^{\circ}38^{'}21^{"}$E) \cite{Monsalve2017, BowmanEDGES2018}. It consists of two zenith pointing single-polarization antennas, a low frequency instrument observing at 50-100 MHz ($13.2 < z < 27.4$), and a high frequency instrument observing at 90-190 MHz ($6.5 < z < 14.8$). The EDGES low-band instrument detected a broad absorption feature in the global \tcm \ signal centered at 78 MHz, with a width of \aprx19 MHz, corresponding to a redshift range of approximately $14 < z < 23$, centered at $z \approx 17$. The depth of the absorption feature was 0.5K, against a background RMS of 0.025K, for a detection signal-to-noise ratio of 37 \cite{BowmanEDGES2018}. An absorption feature was an expected effect of ionizing radiation from Cosmic Dawn. However, the timing and narrowness of the EDGES signal suggests a more rapidly evolving star formation rate than expected \cite{MirochaFurlanetto2019}. Additionally, the amplitude of the absorption is unexpectedly large, signifying a large temperature contrast between the background radiation temperature and the neutral hydrogen spin temperature. Achieving such a large contrast would require at least one of the following revisions to our theoretical models. One possibility is the existence of a population of previously undetected high-redshift radio sources. This would increase the temperature contrast by boosting the background radiation temperature, essentially replacing $T_\gamma$ in Equation \eqref{eq:Tb_theory} with some higher temperature \citep{Feng2018,Sharma2018,EwallWice2018RadioBackground,EwallWice2019RadioBackground,Jana2019,Fialkov2019globalsig}. Alternatively, there is the exciting possibility of exotic physics at play that allows the spin temperature to cool below the temperature expected for a gas that cools adiabatically with our Universe's expansion \citep{Falkowski2018,SlatyerWu2018,HiranoBromm2018,Barkana2018,Costa2018,Moroi2018,Berlin2018,Munoz2018a,Safarzadeh2018,Schneider2018,KovetzFuzzy2018,Kovetz2018,Clark2018,Hektor2018,Mitridate2018,Yoshiura2018,Munoz2018b,cheung2019,Chianese2019,Jia2019,Lawson2019}.

Of course, it is possible that the EDGES anomaly is due to instrumental systematics \citep{BradleyPlane2018} and subtleties in the analysis methods \citep{Hills2018,EDGESreply2018,SinghSubrahmanyan2019,Jana2019,2020MNRAS.492...22S}. Fortunately, other experiments are being built and/or are taking data to confirm or refute EDGES. One such experiment is the Probing Radio Intensity at high-Z from Marion (PRI$^{\mathrm{Z}}$M), a global EoR \tcm \ instrument operating from Marion Island, the larger of the two Prince Edward Islands, located in the Subantarctic Indian Ocean ($46^{\circ}53^{'}13^{"}$S, $37^{\circ}49^{'}11^{"}$E) \cite{Philip2019}. The location was selected for the pristine low frequency observing conditions, due to the remoteness of the location, and to improved ionospheric transparency at low frequencies at the subantarctic latitudes. PRI$^{\mathrm{Z}}$M consists of two dual-polarization "hibiscus" four-square antennas, with bands centered at 70 MHz and 100 MHz, for full observing band of 50-130 MHz ($ 9.9 < z < 27.4$). Another experiment is the Large-aperture Experiment to Detect the Dark Age (LEDA, \cite{Price2018,2021MNRAS.505.1575S}). LEDA is located at the Owens Valley Radio Observatory (OVRO). It is co-located with the OVRO Long Wavelength Array (LWA, \cite{Eastwood2019Pspec,2021MNRAS.506.5802G}) and consists of five LWA antennas outfitted for precision radiometry over the 30 to 85 MHz band. Their location enables LEDA to take advantage of interferometric cross-correlation opportunities with the OVRO-LWA for calibration purposes. A slightly different approach is taken by the Radio Experiment for the Analysis of
Cosmic Hydrogen (REACH; \cite{EloyREACH,2022JAI....1150001C}), which is designed to operate from 55 MHz to $\sim$130 MHz. REACH incorporates a fully Bayesian pipeline that simultaneously models instrumental systematics, foreground uncertainties, and the cosmological signal of interest \cite{2022MNRAS.509.4679A}.

A particularly noteworthy recent effort to confirm or refute EDGES is the Shaped Antenna measured of the background RAdio Spectrum (SARAS; \cite{SinghSARAS_overview2018}) experiment. The most recent version of this experiment, SARAS 3, is a monocone antenna that operates from 43.75 to 87.5 MHz. It is unique in its deployment on a lake that serves as a homogeneous high permittivity medium that takes the place of a ground plane. The latest SARAS measurements are inconsistent with the EDGES signal, disfavouring the interpretation of the EDGES anomaly as one that is due to new physics \cite{2022NatAs.tmp...47S}.

\subsection{EoR Interferometers}

The Hydrogen Epoch of Reionization Array (HERA) is a radio interferometer located at the South African Radio Astronomy Observatory (SARAO) Square Kilometer Array (SKA) site in the Karoo region of South Africa ($30^{\circ}43^{'}17^{"}$S, $21^{\circ}25^{'}41^{"}$E) \cite{2017PASP..129d5001D}. It consists of 350 zenith pointing parabolic dishes 14m in diameter, with focal ratio $f/D = 0.32$. HERA observes at 50-225MHz, corresponding to a redshift range of $5.3 < z < 27$, to probe the intergalactic medium during the cosmic epoch of reionization, when the first energetic objects in the Universe were ionizing the IGM. HERA builds on previous EoR arrays such as the Precision Array Probing the Epoch of Reionization (PAPER) \cite{2010AJ....139.1468P} as well as EoR experiments using existing single dish radio telescopes such as the Giant Metrewave Radio Telescope (GMRT) \cite{10.1111/j.1365-2966.2011.18208.x}. A chief difficulty in previous surveys has been achieving the instrument sensitivities required for a detection of the EoR 21\,cm \ signal, while suppressing foregrounds which are 5-6 orders of magnitude brighter than the signal of interest. HERA brings significantly increased sensitivity to this problem, specifically on angular and spectral scales where the EoR signal may dominate over foregrounds. Indeed, recent upper limits on the power spectrum from HERA \cite{HERAlimits2022} have already put pressure on models of the EoR by disfavouring low levels of X-ray emission from first-generation galaxies \cite{HERA2022science}. As HERA enters a regime of routine science observations with the full array and their full band, direct constraints on redshifts relevant to the EDGES anomaly will be possible in addition to other rich science opportunities that come with directly observing a previously unexplored redshift range at high sensitivity \cite{2017PASP..129d5001D}.

The Murchison Widefield Array (MWA) is a custom-built interferometer located in the Shire of Murchison in the Western Australian desert. It is designed for a wide variety of science cases (including Milky Way science, extragalactic studies, solar physics, and radio transients), but with Cosmic Dawn and Epoch of Reionization studies as key motivators \citep{Bowman2013MWA}. Each element of the interferometer consists of a square $4 \times 4$ grid of crossed dipole antennas (forming a \emph{tile}) whose signals are combined electronically. By changing the relative phases with which the signals from each dipole are combined in a process known as \emph{beamforming}, each tile can be electronically ``pointed" at different parts of the sky. Each tile then serves as a single element in a 256-element interferometer operating between $70$ and $300\,\textrm{MHz}$ \citep{Tingay2013MWA}. At any given instant, $30\,\textrm{MHz}$ of bandwidth and 128 elements can be correlated. True to its broad science case, the tile layout of the MWA is a hybrid between a spread out non-regular configuration and two closely packed, hexagonal grids of 36 tiles each \citep{Wayth2018MWAPhaseII}. The MWA has published a number of upper limits over a broad range of redshifts in recent years ranging from Cosmic Dawn redshifts to reionization redshifts \citep{Dillon2014MWALimits,Dillon2015EmpCov,EwallWice2016MWALimits,Beardsley2016MWALimit,Trott2016CHIPS,Jacobs2016comparison,Trott2019b,Trott2019a,Barry2019}.

The LOw Frequency Array (LOFAR) is a low-frequency radio interferometer with a dense core of elements centred in the Netherlands complemented by a set of remote international elements \citep{LOFAR}. Each interferometric element is known as a station, and is in fact made of two different types of antennas: a set of high-band antennas (HBAs) cover the $110$ to $250\,\textrm{MHz}$ frequency range and a set of low-band antennas (LBAs) cover the $10$ to $90\,\textrm{MHz}$ range. The number of antennas in each station varies from location to location, but the stations in the Netherlands (which give rise to the shortest baselines and therefore are the ones with greatest sensitivity to cosmology) each have 96 LBAs and 48 HBAs. Whereas each LBA is an individual dipole antenna, each HBA is in fact a tile of 16 antennas (in a similar fashion to the MWA) tied together by an analog beamformer. At each station, the outputs from the LBAs and the HBAs are digitized and then digitally beamformed before the data from all the stations are sent to a central correlator located at the University of Groningen. LOFAR is a multi-purpose observatory that accommodates a wide variety of key science projects, including deep extragalactic surveys, transient phenomena, cosmic rays, solar science, cosmic magnetism, and cosmology. It has recently placed upper limits on both Cosmic Dawn \citep{Gehlot2018Limit} and the EoR \citep{Patil2017}.

The Owens Valley Long Wavelength Array (OVRO-LWA) consists of $288$ dipoles, with 251 of these dipoles located within a $200\,\textrm{m}$-diameter compact core and the remaining $32$ dipoles located farther away to provide baselines of up to $1.5\,\textrm{km}$ in length. The elements in the core are arranged in a pseudo-random fashion, enabling a better point-spread function for imaging. This makes OVRO-LWA a powerful instrument for transient science, for instance in its searches for radio emission from gamma ray bursts \citep{Anderson2018GRBs}, gravitational wave events \citep{OVROLWA2019GW}, or from exoplanets \citep{Anderson2017Exoplanets}. With observations spanning an instantaneous bandwidth from $27\,\textrm{MHz}$ to $85\,\textrm{MHz}$, OVRO-LWA has also produced new results for \tcm cosmology, including a detailed set of low-frequency foreground maps \citep{Eastwood2018} and a upper limit on the \tcm power spectrum at $z \approx 18.4$ \citep{Eastwood2019Pspec} and $z \approx 28$ \cite{2021MNRAS.506.5802G}.

\subsection{Low Redshift Interferometers}
Until recently, detections of 21\,cm emission from galaxies were primarily limited to measurements from single-dish telescopes, across regions of the sky that were too small for any cosmological parameter inference \cite{}. Interferometers designed specifically for this detection have since been deployed (CHIME) or are in design phases (CHORD, HIRAX). [is this how we want to say this? The detections have come from mostly GBT, GMRT, ...]

The Canadian Hydrogen Intensity Mapping Experiment (CHIME) is a new cylindrical transit radio interferometer located at the Dominion Radio Astrophysical Observatory ($49^{\circ}19^{'}15^{"}$N $119^{\circ}37^{'}26^{"}$W)\cite{CHIME_instrument_2022}. CHIME operates between 400--800\,MHz in 1024 channels and
consists 1024 dual-polarization feeds spaced equally across
four parabolic $f/0.25$ cylinders (20\,m wide$\times$100\,m long). It is a transit telescope and the cylindrical design provides coverage of $\sim$60\% of the sky over the course of a day. The experiment published recent results showing a positive and significant cross-correlation signal when stacking on galaxy and quasar locations\cite{CHIMEresults2022}, and also featured a novel measurement of the beam pattern using solar measurements\cite{CHIME_sun_2022}. Although CHIME's design was driven by the science requirements for an improved measurement of Dark Energy, it has been a revolutionary new telescope for studies of radio transients, in particular fast radio bursts (FRBs). This has included the first detection within this radio band\cite{2018ATel11901....1B,2019Natur.566..230C}, measurements of repeating sources\cite{2019Natur.566..235C,2019ApJ...882L..18J,2019ApJ...885L..24C,2020ApJ...891L...6F}, a measurement tying FRB emission to a galactic magnetar\cite{2020Natur.587...54C}, and the first measurement of periodic activity in an FRB\cite{2020Natur.582..351C}. CHIME is extending the initial array to include CHORD\cite{2019clrp.2020...28V}, a set of `outrigger' cylinders and dishes at a wider redshift range for improved FRB localization as well as targeting improved dark energy constraints and small scale features in the matter power spectrum which are sensitive to the scalar spectral index ($n_{s}$) and the neutrino mass sum.  

The Hydrogen Intensity and Real-time Analysis eXperiment (HIRAX) is an interferometric radio telescope array which will be located at the SARAO SKA site in the Karoo  ($30^{\circ}41^{'}47^{"}$S, $21^{\circ}34^{'}20^{"}$E) \cite{2016SPIE.9906E..5XN, 2021SPIE11445E..5OS, 10.1117/1.JATIS.8.1.011019}. The full HIRAX array will consist of 1024 parabolic dishes 6m in diameter, for a total collecting area of 28,000 m$^2$. It will observe between 400-800\,MHz, corresponding to a redshift range of $0.8 < z < 2.5$. HIRAX is also a transit telescope. HIRAX uses deep parabolic dishes with a focal ratio of $f/D = 0.23$ to reduce crosstalk between elements in the close-packed array, and to facilitate the calibration of the instrument beams, both of which are proving to be limiting systematics in other interferometric radio arrays. The array will be repointed every six months, building up a map of \aprx 15,000 square degrees of the southern sky, complementary to the sky coverage of the CHIME, and upgraded CHORD, array. The statistical power of this survey is expected to place $7\%$ on the dark energy equation of state parameter, when combined with the measurements from the \textit{Planck} satellite. It is also expected to detect similar numbers of FRBs to the CHIME instrument, given the similar number of beams and total collecting area. Outrigger arrays of \aprx 10 dishes are planned across southern Africa, with baselines on the order of 1000km, to provide localizations of FRBs with an angular resolution of 0.1 arcseconds. 

\subsection{PUMA}
\label{subsec:PUMA}

The Packed Ultra-wideband Mapping Array (PUMA) \cite{2018arXiv181009572C, PUMAAPC} is a proposed next-generation 21\,cm \ intensity mapping array which is optimized for cosmology in the post-reionization era, and which would be well positioned to address other science goals including FRBs and radio transients, pulsar monitoring, non-Gaussianity of the primordial matter power spectrum, inflationary relic features, and multi-messenger astronomy \cite{PUMAAPC}. The reference design calls for PUMA to consist of a hexagonal close-packed array of 32,000 parabolic dishes 6m in diameter, observing at 200-1100\,MHz, corresponding to a redshift range of $0.3 < z < 6$. Due to the essentially modular nature of interferometer arrays, a 5,000 element PUMA-5K array is also considered as a first stage, which would address the same scientific goals but with reduced sensitivity and constraining power, and would naturally build out to the full PUMA-32K array.

PUMA will be capable of addressing scientific questions including (A) the expansion history of the universe, (B) the growth of cosmic structure, (C) primordial non-Gaussianity, (D) relic inflationary features, (E) FRB detection, and (F) pulsar monitoring.

For science goal (A) the PUMA survey will be roughly comparable to a spectroscopic survey of $\aprx 2.5 \times 10^9$ galaxies over the same redshift range, and will produce a nearly sample variance limited measurement of the BAO structure out to a redshift of $z=6$. Additionally, PUMA will be complementary with optical surveys measuring the LSS in the same redshift range in that it will have significantly different instrumental systematics and observational biases, as is discussed in Section \ref{sec:LowzHI}. In particular, the use of \tcm \ emissions from neutral hydrogen as a tracer of large scale structure provides a radically different window of the BAO structure, and the underlying dark matter distribution, from optical galaxies. 

Structure formation in the universe (B) is a balance between the expansion of the universe (driven in late times by dark energy), and gravity (primarily sourced by dark matter) concentrating matter. Precise measurements of the shape of the BAO features provide a sensitive test of General Relativity, and measurements of the basic constituents of the Universe. PUMA will be able to measure the rate-of-growth of the large scale structure with precision roughly an order of magnitude better than current generation surveys, and out to significantly greater redshift, and will place constraints on the sum of the neutrino masses roughly a factor of two stronger than next-generation CMB surveys such as CMB-S4.

In addition to measurements of the evolution of the universe at late times, the large cosmological volume accessible to PUMA would enable constraints to be placed on the mechanisms that generated primordial fluctuations in the matter distribution of the universe during the inflationary epoch. Primordial non-Gaussianity (C) is one of the few observable effects of these primordial fluctuations, reaching $\sigma(f_{NL}) \aprx 1$ would have significant impacts on theoretical models of the primordial universe. PUMA can easily reach this threshold, even given pessimistic assumptions for foreground contamination, in contrast with CMB experiments, for which this threshold is much more difficult to achieve, given the significantly smaller number of primordial modes accessible through the CMB.

PUMA will also be able to place constraints on features in the two-point correlation function (D), constraining the amplitude of linearly-spaced oscillatory features by $5-10\times$ more strongly than current and near-term surveys. PUMA-5K would be comparable to a spectroscopic follow-up survey of LSST galaxies, and the full PUMA-32K survey would approach cosmic variance limits.

FRB (E) detection rates are proportional to the number of synthesized beams in an interferometric array, and to total collecting area. The full PUMA array would therefore be expected to detect FRBS at a rate over two orders of magnitude higher than arrays such as CHIME/CHORD and HIRAX. PUMA-5K would detect \aprx215FRBs per day, and the PUMA-32K \aprx3,500 per day, or two every minute.

Pulsar monitoring (F), and the associated tests of General Relativity would also be an incredible strength of PUMA. It is expected that the SKA1-LOW and SKA1-MID arrays will detect over 10,000 pulsars, but no current telescope facility, including even the SKA, would have enough sky coverage to follow up these pulsars  on a regular cadence. PUMA would truly shine here, given it's large sky coverage, and large collecting area. In fact, PUMA would have significantly larger collecting area than the currently planned generations of the SKA. SKA1-MID is 0.033 km$^2$, and SKA1-LOW is 0.1 km$^2$, while PUMA-5K is 0.14 km$^2$, and PUMA-32K is 0.9 km$^2$. Estimates indicate that PUMA-32K would be capable of observing \aprx10\% of SKA pulsars daily, and 50\% over the course of the nominal survey. Moreover, this collecting area, and therefore pulsar monitoring rate, comes at a fraction of the cost of the SKA, due to the cost savings of smaller dishes (element cost is roughly proportional to the diameter of the primary element raised to the power of 2.7 \cite{stepp2003}), un-powered drift-scan observations, and room temperature receivers. Modeling from the Astro 2020 review estimates that the construction and commissioning of PUMA-5K and PUMA-32K would cost \$59 million and \$373 million in FY2019 USD, respectively \cite{2020arXiv200205072C}.

\section{Conclusions}

We have presented the physics of the 21\,cm emission in the Universe as well as on overview of the current experimentation efforts and proposed future projects. We have also outlined the required technical program to advance the current state of the art which will be a necessary pre-requisite to develop for a thriving experimental program.

We advocate that DOE takes a long-term view of 21\,cm cosmology as a potential growth area over the next few decades. Our recommendations are summarized in the Executive summary at the beginning of this White Paper.

\bibliographystyle{unsrt}
\bibliography{main.bib}

\end{document}